\documentclass[sigconf]{acmart}
\usepackage{amsmath,epsfig}

\long\def\comments#1{}

\usepackage{color}
\usepackage{graphicx}
\usepackage{xspace}
\usepackage{subfigure}
\usepackage{amsmath}
\usepackage{float}
\usepackage{amssymb}
\usepackage{latexsym}
\usepackage{graphicx}
\usepackage{url}
\usepackage{epsfig}
\usepackage{mathrsfs}
\usepackage{multirow}
\usepackage{color}
\usepackage{epstopdf}
\usepackage{listings}
\usepackage[nounderscore]{syntax}
\usepackage{setspace}
\usepackage{framed}
\usepackage{natbib}
\usepackage{mathtools}
\usepackage{algorithm}
\usepackage{algorithmic}
\usepackage{tikz}
\usepackage{fancybox}
\usepackage{eucal} 
\usepackage{pifont}
\usepackage{color}

\usetikzlibrary{arrows, automata}

\newcounter{example}[section]
\renewcommand{\theexample}{\nthesection.\arabic{example}}
\newenvironment{example}{
     \refstepcounter{example}
     {\vspace{1ex} \noindent\bf  Example  \theexample:}}{
     \eop\vspace{1ex}} 

\newcounter{remark}[section]
\renewcommand{\theremark}{\nthesection.\arabic{remark}}

\newcommand{\nthesection}{\arabic{section}}
\newtheorem{property}{Property}

\newcommand{\eop}{\hspace*{\fill}\mbox{$\Box$}}

\newcommand{\stitle}[1]{\vspace{1ex} \noindent{\bf #1}}

\newcommand{\kw}[1]{{\ensuremath {\mathsf{#1}}}\xspace}

\newcommand{\beqn}{\begin{eqnarray*}}
\newcommand{\eeqn}{\end{eqnarray*}}

\newcounter{ccc}

\newcommand{\DeepWalk}{{\sl DeepWalk}\xspace}
\newcommand{\NodetoVec}{{\sl Node2Vec}\xspace}
\newcommand{\DeepSets}{{\sl DeepSets}\xspace}
\newcommand{\StructtoVec}{{\sl Struct2Vec}\xspace}
\newcommand{\MNMF}{{\sl M-NMF}\xspace}
\newcommand{\Gorder}{{\sl GO}\xspace}
\newcommand{\BaseNet}{{\sl DON}\xspace}
\newcommand{\BaseNetPlus}{{\sl DON-RL}\xspace}
\newcommand{\eat}[1]{}
\newcommand{\SimFunc}{\mathtt{S}}
\newcommand{\SlashBurn}{{\sl SlashBurn} \xspace}
\newcommand{\Random}{{\sl Random} \xspace}
\newcommand{\Greedy}{{\sl Greedy} \xspace}
\newcommand{\NE}{{\sl NE}\xspace}
\newcommand{\SNE}{{\sl SNE}\xspace}
\newcommand{\ERGraph}{{Erd{\"o}s-R{\'e}nyi}\xspace}
\newcommand{\Degree}{{\sl Degree} \xspace}
\newcommand{\QFunc}{{\mathcal{Q}(S, v)}}
\newcommand{\QVal}{{\mathcal{Q}}}
\newcommand{\MLQVal}{{{\mathcal{\hat{Q}}}}}
\newcommand{\QStar}{{{\mathcal{Q^*}}}}
\newcommand{\QNet}{{{\mathcal{\hat{Q}}}(S, {\bf v};\Theta)}}
\newcommand{\prob}{{\kw {prob}}}
\newcommand{\COST}{{\kw {cost}}}

\newcommand{\CASE}[1]{\STATE \textbf{case} #1\textbf{:} \begin{ALC@g}}
\newcommand{\ENDCASE}{\end{ALC@g}}

\newcommand{\DEFAULT}{\STATE \textbf{default:} \begin{ALC@g}}
\newcommand{\ENDDEFAULT}{\end{ALC@g}}
\newcommand{\DEFAULTLINE}[1]{\STATE \textbf{default:} }
\newcommand{\Dev}{\kw{E}}
\newcommand{\REINFORCE}{\kw{REINFORCE}}

\title{Graph Ordering: Towards the Optimal by Learning}

\author{Kangfei Zhao}
\email{kfzhao@se.cuhk.edu.hk}
\affiliation{%
  \institution{The Chinese University of Hong Kong}
}
\author{Yu Rong}
\email{yu.rong@hotmail.com}
\affiliation{%
  \institution{Tencent AI Lab}
}
\author{Jeffrey Xu Yu}
\email{yu@se.cuhk.edu.hk}
\affiliation{%
  \institution{The Chinese University of Hong Kong}
}

\author{Junzhou Huang}
\email{joehhuang@tencent.com}
\affiliation{%
  \institution{Tencent AI Lab}
}

\author{Hao Zhang}
\email{hzhang@se.cuhk.edu.hk}
\affiliation{%
  \institution{The Chinese University of Hong Kong}
}

\begin{document}
\def\thepage{\arabic{page}}
\pagestyle{plain}


\clubpenalty=10000 
\widowpenalty = 10000

\begin{abstract}
Graph representation learning has achieved a remarkable success in many graph-based applications,  
such as node classification, link prediction, and community detection. 
These models are usually designed to preserve the vertex information at different granularity and reduce the problems in discrete space to some machine learning tasks in continuous space. 
However, regardless of the fruitful progress, for some kind of graph applications, such as graph compression and edge partition, 
it is very hard to reduce them to some graph representation learning tasks. Moreover, these problems are closely related to reformulating a global layout for a specific graph, which is an important NP-hard combinatorial optimization problem: graph ordering. 
In this paper, we propose to attack the graph ordering problem behind such applications by a novel learning approach. Distinguished from greedy algorithms based on predefined heuristics, we propose a neural network model: Deep Order Network (\BaseNet) to capture the hidden locality structure from partial vertex order sets. Supervised by sampled partial order, \BaseNet has the ability to infer unseen combinations. Furthermore, to alleviate the combinatorial explosion in the training space of \BaseNet and make the efficient partial vertex order sampling , we employ a reinforcement learning model: the Policy Network, to adjust the partial order sampling probabilities during the training phase of \BaseNet automatically. To this end, the Policy Network can improve the training efficiency and guide \BaseNet to evolve towards a more effective model automatically. The training of two networks is performed interactively and the whole framework is called \BaseNetPlus. Comprehensive experiments on both synthetic and real data validate that \BaseNetPlus outperforms the current state-of-the-art heuristic algorithm consistently.  Two case studies on graph compression and edge partitioning demonstrate the potential power of \BaseNetPlus in real applications. 


\end{abstract}

\maketitle

\thispagestyle{plain}

\section{Introduction}

Graphs have been widely used to model complex relations among data,
and have been used to support many large real applications including
social networks, biological networks, citation networks and road
networks.
In the literature, in addition to the graph algorithms and graph
processing systems designed and developed, graph representation
learning have also been studied recently
\cite{DBLP:conf/www/AhmedSNJS13, DBLP:conf/nips/BelkinN01,
  fu2012graph,DBLP:conf/kdd/PerozziAS14,DBLP:conf/kdd/GroverL16,DBLP:conf/www/TangQWZYM15,
  DBLP:conf/www/TsitsulinMKM18,
  DBLP:conf/kdd/ZhangCWPY018,DBLP:conf/aaai/WangCWP0Y17,DBLP:conf/icde/FengCKLLC18,DBLP:conf/icde/HuAMH16,DBLP:conf/ijcai/ManSLJC16}.  They are
designed to preserve vertex information at different granularity from
microscopic neighborhood structure to macroscopic community structure,
and are effectively used for node classification, link prediction,
influence diffusion prediction, anomaly detection, network alignment,
and recommendation. The surveys can be found in \cite{cui2018survey,
  DBLP:journals/debu/HamiltonYL17}.
  
\begin{figure}[t]
\centering
\subfigure[Order by {\sl GO}~\cite{DBLP:conf/sigmod/WeiYLL16}]{%
 \includegraphics[width=0.7\columnwidth]{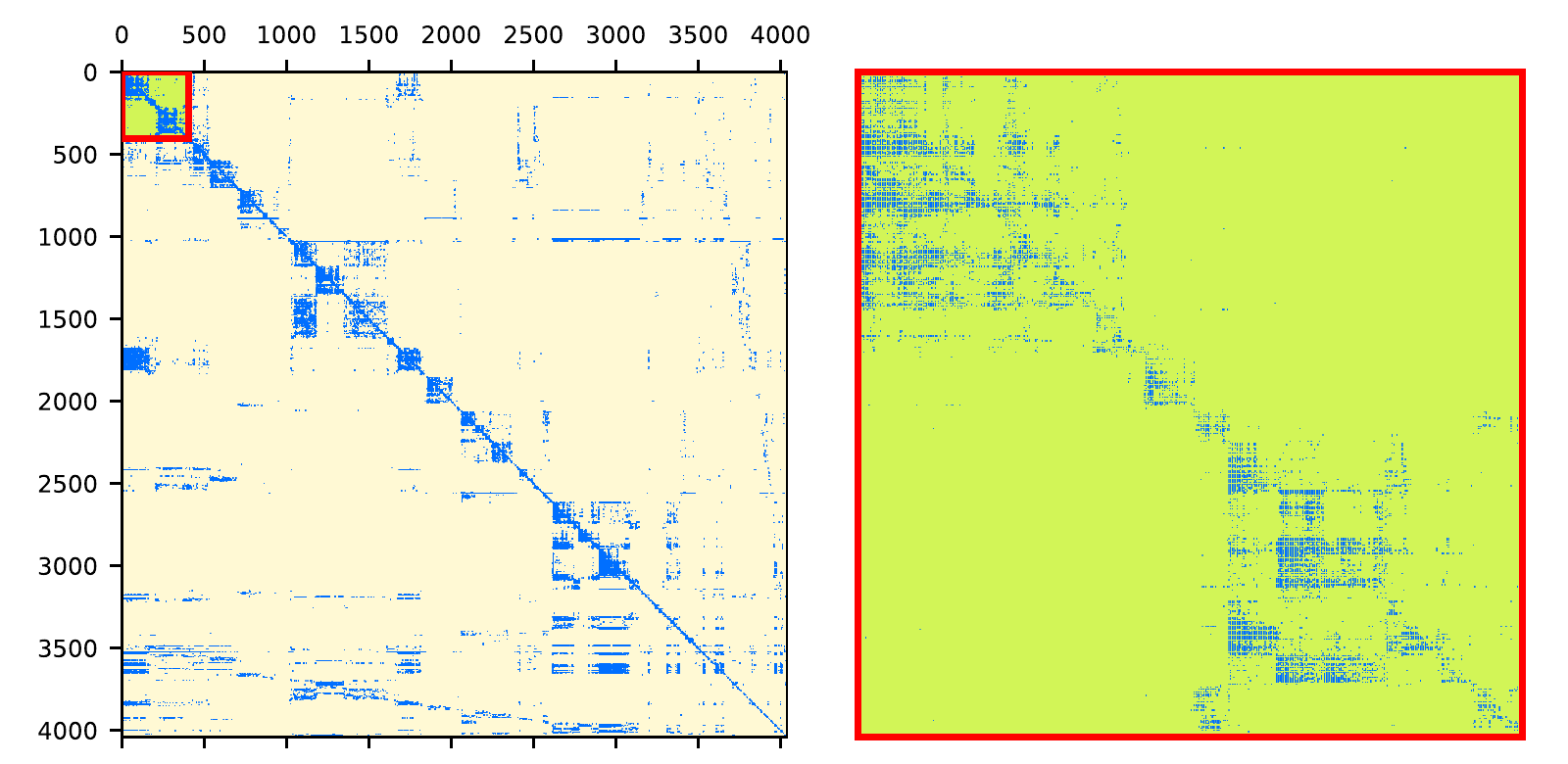}
 \vspace*{-0.2cm} 
  \label{fig:orderbyalg}}\\
\subfigure[Order by \BaseNetPlus]{%
  \includegraphics[width=0.7\columnwidth]{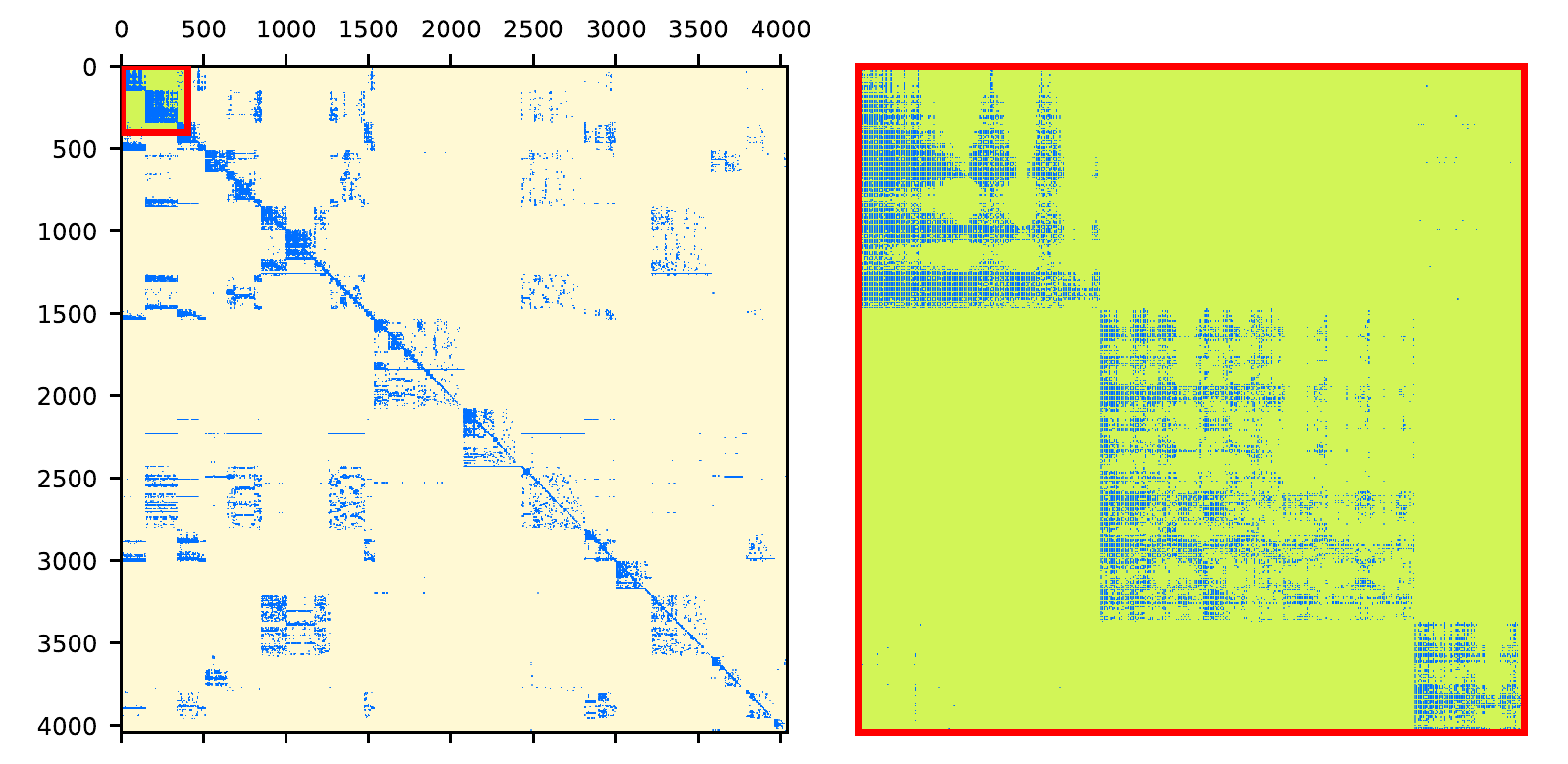}
 \vspace*{-0.2cm}   
  \label{fig:orderbyml}}
\vspace*{-0.2cm}
\Description{Matrix Visualization for the Reordered Facebook}
\caption{Matrix Visualization for the Reordered Facebook}
\vspace*{-0.2cm}
\label{fig:matrixvisual}
\end{figure}

In this paper, we concentrate ourselves on a rather different set of
graph problems: graph
visualization~\cite{DBLP:journals/cgf/BehrischBRSF16}, graph
compression~\cite{KangF11}, graph
edge partitioning~\cite{DBLP:conf/kdd/BourseLV14,
  DBLP:conf/kdd/ZhangWLTL17}.
%
%
Here, graph visualization is to provide an effective way to visualize
complex data, graph compression is to find a compact edge layout,
and graph edge partitioning is to
partition edges instead of vertices for better load
balancing~\cite{graphlab, DBLP:conf/osdi/GonzalezXDCFS14}.  Such
problems can be addressed if we can arrange all vertices in a good
order.
The-state-of-art graph ordering~\cite{DBLP:conf/sigmod/WeiYLL16}
reassigns every vertex a unique number in $[1..n]$ where $n$ is the
number of vertices in the graph by maximizing a cumulated locality
score function for vertices in a sliding window of size $w~(> 1)$,
which is an NP-hard combinatorial optimization problem as a general
form of the maximum Travel Salesman Problem~(TSP) ($w = 1$).  The {\sl
  GO} algorithm proposed in~\cite{DBLP:conf/sigmod/WeiYLL16} achieves
$\frac{1}{2w}$-approximation, 
and its real performance is very close to the optimal in their testing, as it is close to the upper bound of the optimal solution.

%
%

Given such a high-quality solution for graph ordering, a natural
question is if we can do better by learning.
The answer is yes.  We
demonstrate it in Fig.~\ref{fig:matrixvisual}, which shows the matrix
visualization for the Facebook~\cite{snapnets} reordered by the {\sl
  GO} algorithm and our new learning-based approach over graph. 
The left figures are the overall visualizations and the right are the zoom-in details of the upper-left part of the overall ordering. 
Comparing with the {\sl GO}
algorithm~\cite{DBLP:conf/sigmod/WeiYLL16}~(Fig.~\ref{fig:orderbyalg}),
our learning-based approach~(Fig.~\ref{fig:orderbyml}) keeps the
compact permutation better from the local perspective, and has the
advantage of avoiding stucking in forming local dense areas from the
global perspective.
Moreover, the learned permutation has a large potential to support other real applications. 
Based on our case studies, the learned permutation reduces the storage costs larger than $2\times$ compared with the order for real-graph compression~\cite{KangF11}. 
And the edge partitioning deriving from our approach outperforms the widely-used partitioning algorithm~\cite{DBLP:conf/kdd/BourseLV14} up to 37\%. 

The main contributions of this paper are summarized below.
\begin{itemize}
    \item For the graph ordering problem, distinguished from 
      the traditional heuristics-based algorithm, 
      we propose a neural network based model: Deep Order
      Network (\BaseNet) to replace the predefined evaluation
      function to capture the complex implicit locality in  
      graphs. Hence our proposed model can produce a better graph ordering 
      than the current state-of-the-art heuristics-based algorithm.  
    \item To address the combinatorial explosion in the training space
      of \BaseNet, we exploit the idea of reinforcement learning and
      construct a policy network to learn the sampling probability of
      vertices. This policy network enables \BaseNet to focus on
      vertices which have more remarkable contribution in the solution
      and guides \BaseNet to evolve towards a more effective model
      automatically. The whole learning framework is called
      \BaseNetPlus. 
    \item We conduct extensive experiments on both synthetic and
      real-world graphs with different properties. Experimental
      results show that \BaseNetPlus consistently outperforms the
      other algorithms. Furthermore, we conduct two case studies:
      graph compression and graph edge partitioning to demonstrate the
      potential value of our algorithm in real applications.  
\end{itemize}

\comments{
First, we
discuss the weakness of traditional heuristics-based greedy algorithm.
Second, we design a supervised learning framework for dealing with
graph ordering problem. The core of this framework, called Order
Network, samples the training data on the graph locally to learn a
graph representation.  The representation clusters vertices intend to
form high-quality partial solutions together in low-dimensional vector
space while keeping those in low-quality solutions away.  Third, to
improve the effectiveness of the model, we design an adaptive training
approach to enabling much training effort to be paid on vertices which
have more remarkable contribution in the solution.  Fourth, we conduct
extensive experimental studies on real and synthetic graphs. Our
model-based approach generates much better vertex permutation with
regards to the optimization objective.  Moreover, Our study indicates
that even simple neural network model has the potential to surpass
traditional greedy algorithms in solving NP optimization problem.
}


The paper is organized as follows. 
Section~\ref{sec:pre} introduces the preliminaries, including the problem definition, the intuitions and challenges. 
Section~\ref{sec:overview} gives an overview of our learning framework, which is composed of a central network for ordering and an auxiliary Policy Network. 
The design principles of these two neural networks are introduced in Section~\ref{sec:base} and Section~\ref{sec:RL}, respectively. 
Section~\ref{sec:exp} presents the experimental result of our approach in solving the graph ordering problem. 
We review the related works in Section~\ref{sec:rw}.
Finally, we conclude our approach in Section~\ref{sec:conclusion}.

\begin{table}[t]
\begin{center}
{\scriptsize
\caption{Frequently Used Notations} 
\label{tab:notation}
\vspace*{-0.3cm}
\begin{tabular}{|p{0.23\columnwidth}|p{0.7\columnwidth}|} \hline
{\bf Notations} & {\bf Definitions} \\ \hline\hline
$G = (V, E)$  & directed graph with the vertex set $V$, the edge set $E$. \\ \hline
$\SimFunc(u, v)$ & The pair-wise similarity function of vertex $u$ and $v$. \\ \hline
$F(\Phi)$ & The locality score function of a permutation $\Phi$. \\ \hline
$w$ & The given window size. \\ \hline
$\QNet$ & Deep Order Network~(\BaseNet). \\ \hline
$\Pi(s;\Theta')$ & Policy Network. \\ \hline
${\bf \prob}$~(${\bf \prob_{t}}$)  & The vertices sampling probability (at time $t$). \\ \hline 
${\bf a}$~(${\bf a_t}$) & The tuning action of vertices sampling probability (at time $t$). \\ \hline
$ D_{t} / D_{\Dev}$ & The training data at time t / The evaluation data. \\ \hline
\end{tabular}
}
\end{center}
\vspace*{-0.3cm}
\end{table}

\begin{algorithm}[t]
{
\caption{{\sl GO}~($G = (V, E)$) }
\label{alg:Greedy}
\begin{algorithmic}[1]
\STATE  $S \leftarrow \emptyset$;
\WHILE {$|S| \neq |V|$}
\STATE $v^* = \arg max_{\forall v \in V} \QFunc$ ;
\STATE $S \leftarrow S \cup \{ v^*\}$;
\ENDWHILE
\RETURN $S$;
\end{algorithmic}
}
\end{algorithm}

\begin{figure}[t]
\centering
\subfigure[A Directed Graph]{%
 \includegraphics[width=0.47\columnwidth]{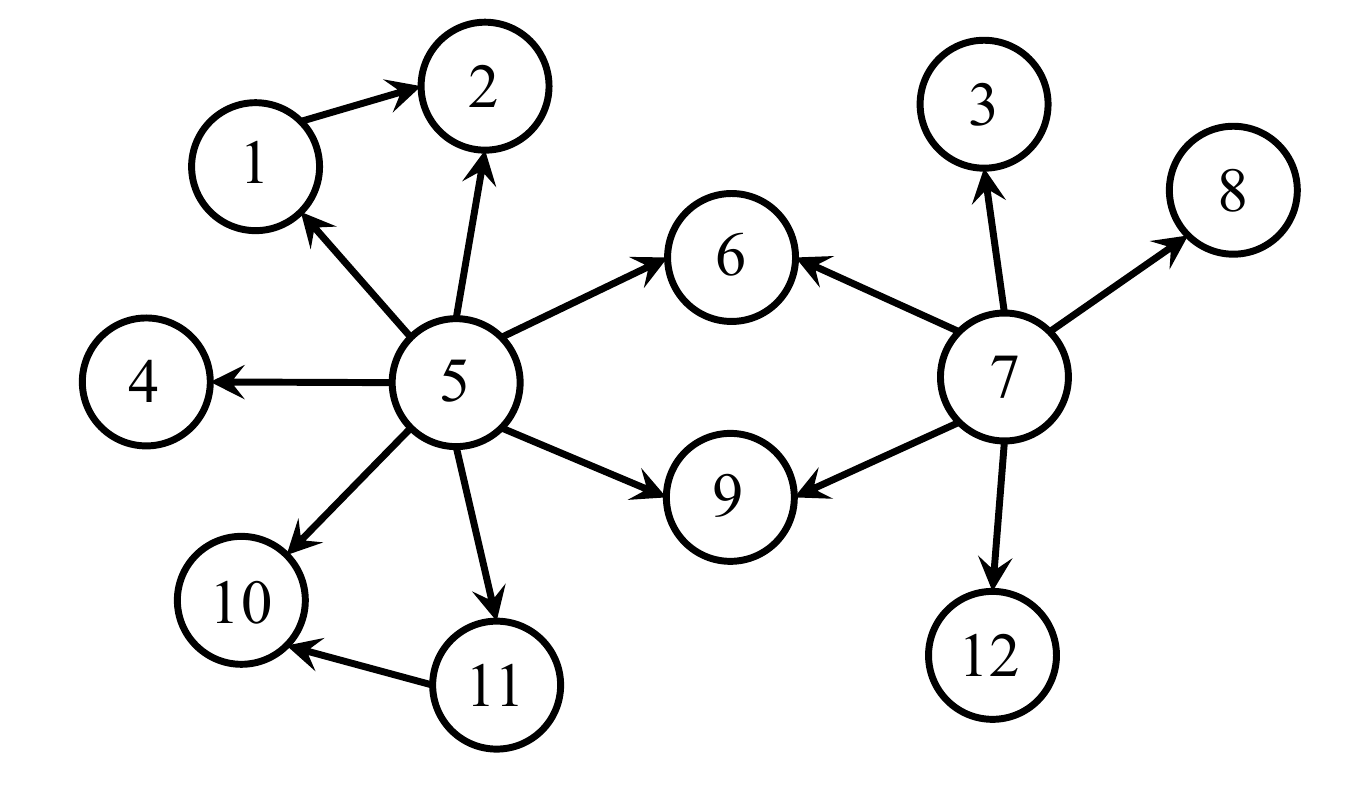} 
  \label{fig:example:1}}
\subfigure[Maximize $F(\Phi)$]{%
  \includegraphics[width=0.47\columnwidth]{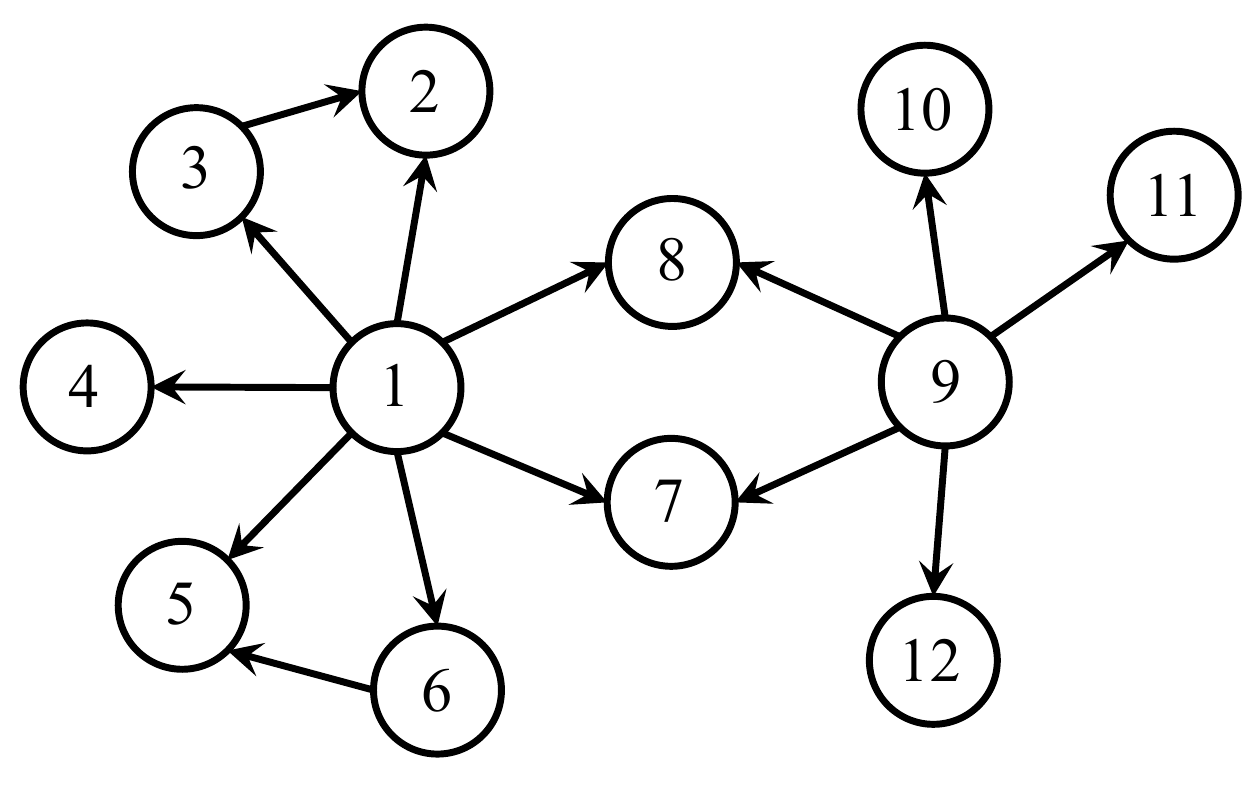}
  \label{fig:example:2}}
\caption{An Example of Graph Ordering~\cite{DBLP:conf/sigmod/WeiYLL16}}
\vspace*{-0.2cm}
\label{fig:example}
\end{figure}

\section{Preliminaries}
In this section, first, the graph ordering problem and its existing solution are introduced. Then, we elaborate on our intuitive improvement and the challenges. 
\label{sec:pre}
\subsection{Graph Ordering}

Given a graph $G = (V, E)$ where $V(G)$ is a set of vertices
and $E(G)$ is a set of edges of $G$, the graph ordering problem is to
\emph{find the optimal order for all vertices}. In brief, let $\Phi(\cdot)$
be a permutation function which assigns a vertex a unique number in
$[1..n]$ where $n = |V(G)|$. The problem is to find the optimal graph
ordering $\Phi(\cdot)$ by maximizing a cumulated locality score
$F(\cdot)$ (Eq.~(\ref{eq:fscore})) over a sliding window of size $w$.
\begin{equation}
F(\Phi)  = \displaystyle{\mathop{\Sigma}_{\substack{0 < \Phi(v) -
      \Phi(u) \leqslant w}}} \SimFunc(u, v)  
\label{eq:fscore}
\end{equation}
Here, $u$, $v$ are two vertices ordered within a window of size $w$ in
the permutation, and $\SimFunc(u, v)$ is a pair-wise similarity
function to measure the closeness of $u$ and $v $.  In
\cite{DBLP:conf/sigmod/WeiYLL16}, $\SimFunc(u, v)$ is defined as
$\SimFunc(u, v) = \SimFunc_s(u, v) + \SimFunc_n(u, v)$, where
$\SimFunc_s(u, v)$ is the number of times that $u$ and $v$ are
sibling,
%
%
and $\SimFunc_n(u, v)$ is the number of times that $u$ and $v$ are
neighbors.
%
%

\begin{example}
Taking the directed graph shown in Fig.~\ref{fig:example:1} as an example, there are 12 nodes numbered from 1 to 12, which is considered
as one possible graph ordering. 
Table~\ref{tab:example:fvalue} lists the pair-wise $\SimFunc(u, v)$ values, in a matrix form for vertices $1-5$ in Fig.~\ref{fig:example:1} due to limited space. 
Given a window size $w = 3$, consider a partial permutation of length 3, $\Phi = [1, 2, 5]$ for the vertices in Fig.~\ref{fig:example:1}, its locality score $F(\Phi) = \SimFunc(1, 2) + \SimFunc(2, 5) + \SimFunc(3, 5) = 4$. 
Fig.~\ref{fig:example:2} shows the optimal graph ordering which maximizes $F(\Phi)$ given the window size $w = 3$. 
\end{example}

\begin{table}[t]
\begin{center}
{\scriptsize
\caption{$\SimFunc(u, v)$ Values for Fig.~\ref{fig:example:1}}
\label{tab:example:fvalue}
\begin{tabular}{|c|c|c|c|c|c|} \hline
{\bf $\SimFunc(u, v)$}  & 1 & 2 & 3 & 4 & 5 \\ \hline 
1 & - & 2 & 0 & 1 & 1 \\ \hline
2 & 2 & - & 0 & 1 & 1 \\ \hline
3 & 0 & 0 & - & 0 & 0 \\ \hline
4 & 1 & 1 & 0 & - & 1 \\ \hline
5 & 1 & 1 & 0 & 1 & - \\ \hline
\end{tabular}
}
\end{center}
\vspace*{-0.3cm}
\end{table}

In general, maximizing $F(\Phi)$ is a variant of maximum TSP (for $w =
1$) with the sum of weights within a window of size $w > 1$ and thus is 
NP-hard as proven in~\cite{DBLP:conf/sigmod/WeiYLL16}.

Wei et al. in ~\cite{DBLP:conf/sigmod/WeiYLL16} proposed a greedy
algorithm (named {\sl GO}), as sketched in Algorithm~\ref{alg:Greedy},
which iteratively extends a partial solution $S$ by a vertex
$v^*$. The vertex $v^*$ is selected by maximizing a potential function
$\QFunc$, which measures the quality of the partial solution extended
by a vertex $v$, based on the current partial solution $S$.
%

%

%
In details, in $i$-th iteration, it computes the cumulated weight
$k(v)$~(Eq.\ref{eq:kscore}) for each vertex $v$ and inserts the one
with maximum $k(v)$ in the permutation:
\begin{equation}
k(v)  = \displaystyle{\mathop{\Sigma}_{\substack{j = max(1, i-w)}}} \SimFunc(v_j, v),
\label{eq:kscore}
\end{equation}
where $k(v)$ serves as the potential function $Q$, $\{ v_j | j =
max(1, i-w) \}$ are the last $max(1, i-w)$ inserted vertices.  This
procedure repeats until all the vertices in $G$ are placed. 
The total time complexity of {\sl GO} is $O(wd_{max}n^2)$, 
where $d_{max}$ denotes the maximum in-degree of the graph $G$. 
There are $n$ iterations, and in each iteration, 
it scans the remaining vertices in $O(n)$ and computes 
the score $k(v)$ for the scanned vertex $v$ in $O(w d_{max})$. 
The algorithm achieves $\frac{1}{2w}$-approximation for maximizing
$F(\Phi)$. 
It can achieve high-quality approximations with small window size 
and its practical performance is close to the optimal in the experiments.

\subsection{Intuitions and Challenges}

\stitle{Intuitions:}
Given the current greedy-based framework, Algorithm~\ref{alg:Greedy} can be 
regarded as a discrete Markov Decision Process (MDP)~\cite{DBLP:books/lib/SuttonB98} , where the states are feasible partial permutations, actions are to insert one vertex to the permutations.  
For traditional greedy algorithm with predefined heuristics, the solution is built by a deterministic trajectory which has locally optimal reward.  
With regards to the graph ordering problem, $Q(s,a)$ (Eq.~(\ref{eq:bellman})) is simply approximated by the evaluation function $\QFunc$, which is specifically defined in Eq.~(\ref{eq:kscore}).
A natural question arises that \emph{can we do better by learning an expressive, parameterized $\QNet$ instead.} 

Inspired by universal 
approximation theorem~\cite{csaji2001approximation}, 
deep reinforcement learning~(DRL) learns a policy $\pi$ from states to actions that specifies what action to take in each state. We can formulate Algorithm~\ref{alg:Greedy} as MDP, which is composed by a tuple $(\mathcal{S}, \mathcal{A}, \mathcal{R}, \mathbb{P}, \gamma)$ : 
1) $\mathcal{S}$ is a set of possible \emph{states}, 
2) $\mathcal{A}$ is a set of possible \emph{actions}, 
3) $\mathcal{R} = r(s_t,a_t)$ is the distribution of \emph{reward} given state-action $(s_t, a_t)$ pair. 
4) $\mathbb{P} = P(s_{t+1}|s_t, a_t)$ is the transition probability given $(s_t, a_t)$ pair. 
5) $\gamma$ is the reward discount factor. 
The \emph{policy} $\pi$ is a function $\pi$: $\mathcal{S} \to \mathcal{A}$ that specifies what action to perform in each state.
%

The objective of RL is to find the optimal policy $\pi^*$ which maximizes the expected cumulative discounted reward. This cumulative discounted reward can be defined by the Bellman Equation as below for all states recursively:
\begin{equation}
Q(s, a)  = \mathbb{E}[ r(s, a) + \gamma max_{a'}Q(s', a')| s, a],
\label{eq:bellman}
\end{equation}
where $Q(s, a)$ and $Q(s', a')$ are the maximum expected cumulative rewards achievable from the state-action pair $(s, a)$ and the previous state-action pair $(s',a')$, respectively.
To avoid exhaustively exploring the MDP, Monte Carlo simulation is used to sample the \emph{state-action trajectory} $(s_0, a_0, r_0, s_1, a_1, r_1, \cdots, s_T, a_T, r_T)$ to estimate the values of downstreaming state-action pairs.  In this vein, DRL is adopted to solve the combinational optimization problem~\cite{DBLP:journals/corr/BelloPLNB16, DBLP:conf/nips/KhalilDZDS17}, whose policy is learned by specific deep neural networks.

\begin{figure*}[t]
{
\centering
  \includegraphics[width=0.68\linewidth]{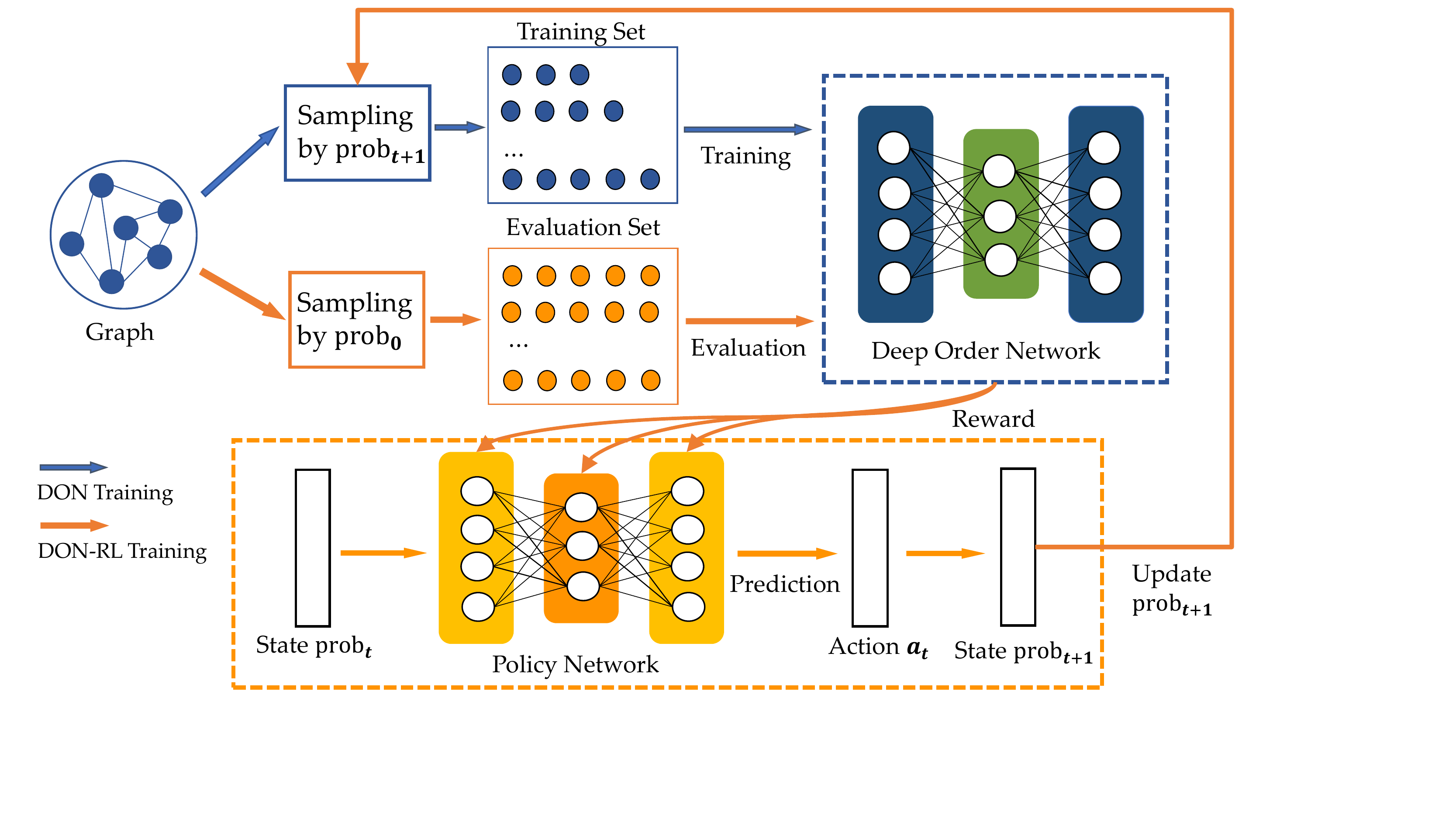}
\vspace*{-0.2cm}
\caption{The overview of \BaseNetPlus. \BaseNetPlus contains two components. One is in the blue box named \BaseNet. The other is in the orange box named the Policy Network. \BaseNetPlus learns a $\QNet$ by sampling partial solutions over a graph. The Policy Network takes the sampling probability ${\bf \prob_t}$ as input to predict a sampling tuning action ${\bf a_t}$. The training of the two networks is performed interactively, where the updating of the Policy Network is controlled by the evaluation result of \BaseNet. }
\label{fig:overview}
}
\vspace*{-0.4cm}
\end{figure*} 

\stitle{Challenges:}\label{sec:chllenges}
Recently, these RL-based approaches~\cite{DBLP:journals/corr/BelloPLNB16, DBLP:conf/nips/KhalilDZDS17} have surpassed traditional greedy strategies on finding better feasible solutions. However, in practice, the pure RL framework is inefficient as the discrete action space becomes large. This drawbacks obstacles RL-base approaches to solve the graph ordering problem over a single large real-world graph. Concretely, RL algorithms use Thompson Sampling~\cite{DBLP:conf/icml/GopalanMM14} to explore trajectories and estimate the expected cumulative reward~(Eq.~(\ref{eq:bellman})).
The larger the action space is, the lower probability trajectories with high accumulated
reward can be sampled in limited steps. The low-quality samples will
extremely degrade the effectiveness of the model as well as the convergence in the training phase.  Therefore,
\cite{DBLP:journals/corr/BelloPLNB16, DBLP:conf/nips/KhalilDZDS17} can only
support training on graphs with tens to one hundred of vertices, where
their optimal solutions are provided by state-of-the-art integer
programming solvers~\cite{cplex201412, applegate2006concorde}. 
Recall that AlphaGo~\cite{DBLP:journals/nature/SilverHMGSDSAPL16} has
$19 \times 19$ actions at most, which in fact is an enormously large
action space requiring high computation power of a large cluster to
explore.  Unfortunately, 
due to the combinatorial explosion in 
the training space of $\QFunc$, any real large graph has a larger searching 
space than that of AlphaGo.

In a nutshell, the construction of the parametric model $\QNet$ and the efficient training in an exponential training space are two main challenges when we try to adopt the model-based framework to solve the graph ordering problem.

\section{A New Approach: \BaseNetPlus}
\label{sec:overview}
\comments{
In this section, we give an overview on  how to devise and evaluate the function $\QFunc$. Inspired from universal approximation theorem \cite{csaji2001approximation}, we also use an expressive, parameterized $\MLQVal$ function instead, learned by a neural network.  However, due to the combinatorial explosion of the input partial solution $S$, it is impractical to memorize all possible partial solutions in $\MLQVal$. To guide $\MLQVal$ to focus on vertices which are important to construct good solutions, we construct the Policy Network to sampling the partial solution during training phase dynamically. 

To this end, to capture the complex combinatorial structure of over a single large graph in solving the graph ordering problem, we propose a new framework: Deep Ordering Network with Reinforcement Learning (\BaseNetPlus). Fig.~\ref{fig:overview} demonstrates the overall learning framework. The core is the parametric model $\QNet$ (at the top of Fig.~\ref{fig:overview}), called Deep Order Network (\BaseNet). Training of \BaseNet is guided by a policy gradient RL model (at the bottom of Fig.~\ref{fig:overview}), called Policy Network.}

In order to address the challenges mentioned in Section~\ref{sec:chllenges}, we propose a new framework: Deep Ordering Network with Reinforcement Learning (\BaseNetPlus) to capture the complex combinatorial structures of a single large graph in solving the graph ordering problem. To make the better approximation of the evaluation function $\QFunc$, we propose a new model: Deep Order Network (\BaseNet) which take the partial vertex order set $S$ as input and predict the next vertex which can maximize the cumulated locality score. Furthermore, to prune the exponential training space and improve the training of \BaseNet efficiency, we employ a policy gradient RL model: the Policy Network to explore the valuable vertex order segments efficiently. Fig.~\ref{fig:overview} demonstrates the overall learning framework.

\stitle{Deep Order Network}: As a parametric permutation-invariant model of the evaluation function $\QFunc$ (Algorithm~\ref{alg:Greedy}), \BaseNet takes a partial vertex order solution, i.e., a set of vertices as input, and outputs the probability distribution of the decision of next vertex to be selected. In other words, \BaseNet is to learn a parameterized $\QNet$ function by exploiting vertex sets over the whole graph $G$, where every set encodes the contribution of vertices in the set for constructing $F(\Phi)$ if they coexist in a window of size $w$. \BaseNet is trained by supervised learning over partial solutions. Here, a partial solution is a set of $m \leq w$ vertices for a window of size $m$ by first sampling $m-1$ vertices from $G$ followed by computing the likelihood of th e last vertex to be in the window. The last vertex to be added for a partial solution is by the ground truth of $\QVal$ value, which is easy to be calculated when $m$ is small. 

Since the function $\QNet$ has the property of permutation invariant regarding any permutation of the vertices in $S$, we employ a \emph{permutation invariant neural network} architecture to model it. We present the design and principle of our Deep Order Network in Section~\ref{sec:base} in details.

%

%

\stitle{RL-based Training}: In practice, the space of the all possible partial vertex order solution is exponential with respect to the number of vertices of a graph. Theoretically, a neural
network can memorize and serve as an oracle for any query solution.
However, it is intractable to enumerate all the partial solutions in
training \BaseNet. Hence, sampling the high-quality partial solutions, i.e., sets of 
vertices, is vital to train \BaseNet model with high generalization ability which can inference unseen vertex combinations rather than memorization.

In other words, \BaseNet should have the ability
to inference unseen vertex combinations rather than memorization. 
Hence, sampling the high-quality partial solutions, i.e., sets of 
vertices, is vital for the training of \BaseNet. 

%
Rather than specifying
the sampling probability uniformly or by human experience,  we employ Policy Network to explore how to make adjustments on the probability,
based on the evaluation result of \BaseNet dynamically. 
In this vein, we regard \BaseNet as an environment
and the Policy Network is an agent which obtains rewards from the
environment and predicts an adjusting action. 
The sampler of vertex sets adopts the adjusted sampling probability (i.e., the state transformed 
from adjusting action of the Policy Network),   
to sample the training data for next time slot and feeds these data to 
\BaseNet.
After \BaseNet is trained for
a certain number of iterations, 
the evaluation result of \BaseNet is received by the Policy Network as the rewards for 
updating its weights, as well as subsequent prediction. 
The training of the two networks is performed interactively as shown
in Fig.~\ref{fig:overview}. 

\comments{
The Policy Network updates the input
pipeline of \BaseNet, after \BaseNet is trained for
a certain number of iterations. 
The evaluation result of \BaseNet is received by the Policy Network for 
updating its parameters, as well as subsequent prediction.
}

The significance of using an RL framework to control the input
pipeline is two folds. First, it improves the training efficiency by a
self-adaptive sampling scheme. Second, it guides \BaseNet to
evolve towards a more effective model automatically.

In contrast to applying RL to solve the problem directly, using RL to make decisions on the sampling 
probability adjustment is feasible. 
On one hand, in practice, the action space of this probability adjustment task is smaller than that of directly attacking the whole problem. 
Since a main property of real-world graph is scale-free, we find that among the vertices there exists skewness regarding the significance to build high-quality solutions so that the pattern of rising up and pushing down vertex sampling probability is not diverse.  
On the other hand, in \BaseNetPlus framework, RL only serves as an auxiliary model to improve the training effectiveness and efficiency of \BaseNet, instead of estimating $\QNet$ directly. 
Thus, the effectiveness of \BaseNet is less sensitive to low-quality state-action trajectories. 
Section~\ref{sec:RL} introduces the RL formulation and our algorithm in
details.

\eat{ 
\stitle{Prediction}: Upon the model converges, we use the same greedy
strategy (Algorithm~\ref{alg:Greedy}) to construct a solution by
expending the permutation. In each iteration (line 3-4), we predict
the probability of a vertex to be with the current partial solution
$S$ (e.g., the past inserted $w-1$ vertices) using the Order Network
constructed, and pick the vertex with the highest probability as the
$v^*$ to be added into $S$. The solutions can be further refined by
advanced search strategies and local search heuristics, which are
beyond the scope of this paper.
}

\section{Deep Order Network}
\label{sec:base}
In this section, we propose the model Deep Order Network~(\BaseNet) as a parameterized evaluation function $\QNet$ to replace the evaluation function $\QFunc$ in Algorithm~\ref{alg:Greedy}. 
Concretely, $\QNet$ takes a vectorized
representation of partial solution $S$ of $w-1$ vertices as input, and outputs a vector
${\bf v} \in \mathbb{R}^N$ where ${\bf v}_i$ represents the likelihood
of inserting a vertex $v_i$ to the permutation, given the
current partial solution $S$. And $\QNet$ being
learned will replace $\QFunc$ seamlessly in
Algorithm~\ref{alg:Greedy} to generate a solution.
%

To learn $\QNet$, we learn a vertex representation with the
preservation of quality of the partial solutions, i.e., partial
permutations $\Phi(S)$ which are composed by any subsets of vertices
of size $w$.
For a given graph, this representation encodes the hidden optimality
structure of vertices in $F(\Phi(S))$ within a window.
%
Intuitively, vertices with a higher $\MLQVal$ value tend to cluster in
the vector space, whereas vertices with a lower $\MLQVal$ value are
kept away.
A high-quality feasible solution can be constructed by set
expansion towards a high $\MLQVal$ value in the vector space.
%
It is worth mentioning that, for a partial permutation $S$ within a window,
the locality score function $F(\Phi(S))$ should permutation invariant with
respect to $\Phi(S)$. Therefore, the learned evaluation function $\MLQVal$ 
is permutation invariant with any permutation of the elements
in $S$.

\begin{property}
\cite{DBLP:conf/nips/ZaheerKRPSS17}~A function of sets $f: X \rightarrow y $, is permutation invariant,
i.e., for any set ${\bf x} = \{ x_{1}, \cdots, x_{M} \} \in X$ with
any permutation $\pi$ of the elements in ${\bf x}$, $f(\{ x_{1},
\cdots, x_{M} \}) = f(\{ x_{\pi(1)}, \cdots, x_{\pi(M)} \})$.
\end{property}

\DeepSets~\cite{DBLP:conf/nips/ZaheerKRPSS17} is a neural network architecture to model the permutation invariant function defined on sets.  For a set {\bf x} whose
domain is the power set of a countable set, any valid, i.e.,
permutation invariant function can be represented in the form of
Eq.~(\ref{eq:deepset}), for two suitable functions $\phi$ and $\rho$.
\begin{equation}
{\small
f({\bf x}) =  \rho ( \Sigma_{i = 1}^M\phi(x_i))
}
\label{eq:deepset}
\end{equation}
In Eq.~(\ref{eq:deepset}), $x_i \in \mathbb{R}^D$ is the $i$-th
element in the set $x$, $\phi: \mathbb{R}^D \rightarrow \mathbb{R}^K$, and
$\rho: \mathbb{R}^K \rightarrow \mathbb{R}$. Intuitively, $\phi$
generates the element-wise representation for each $x_i$, and $\Sigma$ is
a pooling function which performs a generalization of classic
aggregation functions, e.g., \kw{SUM}, \kw{AVG} and \kw{MAX} on the
set.  The added-up representation is processed by $\rho$ in the
same manner as any neural network.
In our problem, the domain of partial solutions $S$ is the combinations of $V$ with size $w-1$, where $V = \{v_1, v_2, \cdots v_N\}$ is the
vertex set. The evaluation function
$\QNet$ for any such $S$ can be learned by parameterizing
the two functions ($\phi$ and $\rho$) with a specified pooling function
$\Sigma$.

We measure the output of $\QNet$ by a distribution $p({\bf v} | S)$, where $p({v_i} | S)$ is the $i$-th element of $p({\bf v} | S)$, meaning how probable the $i$-th vertex can be added into the
current partial solution $S$.  With the Bayes rule, this probability
is in Eq.~(\ref{eq:output}).
\begin{equation}
{\small
p(v_i |S) = \frac{p(v_i, S)}{p(S)} \propto p(S \cup \{v_i\}) 
}
\label{eq:output}
\end{equation}
In Eq.~(\ref{eq:output}), $p(S \cup \{v_i\})$ is the marginal
probability of the extended solution, which can be estimated by
computing $F$ score of the partial permutation $\phi(S \cup \{v_i\})$ explicitly.
\begin{equation}
{\small
p(S \cup \{v_i\}) 
= \frac{F(\phi(S \cup \{v_i\}))}{ \Sigma_{i=1}^N F(\phi(S \cup \{v_i\}))}
}
\label{eq:estimate}
\end{equation}
Additionally, we use the normalized $F(\phi(S \cup \{v_i\}))$ values ~(Eq.~(\ref{eq:estimate})) to make training more
stable.  
As shown in Fig.~\ref{fig:overview}, a training instance is
constructed by sampling a partial solution $S$ with $w-1$ vertices as input feature, 
then $p({\bf  v} | S)$ (Eq.~(\ref{eq:estimate})) serves as the soft label for training.  We
use the cross entropy as the training loss, as given in
Eq.~(\ref{eq:lossfunc}).  Here, $\hat{p}({\bf v} | S)$ is the prediction of \BaseNet.
\begin{equation}
{\small
L(S) = \Sigma_{i=1}^{N} - p({v_i} | S) \log(\hat{p}({v_i} | S)) 
}
\label{eq:lossfunc}
\end{equation}
To model $\QNet$, batches of partial solutions are fed
into \BaseNet, where a partial solution $S$ is generated by
sampling $w - 1$ vertices over $V$ by a probability distribution $\prob$ without
replacement.
%
%
%
\begin{figure}[t]
\centering
\subfigure[Wiki-vote]{%
 \includegraphics[width=0.475\columnwidth,height=3.1cm]{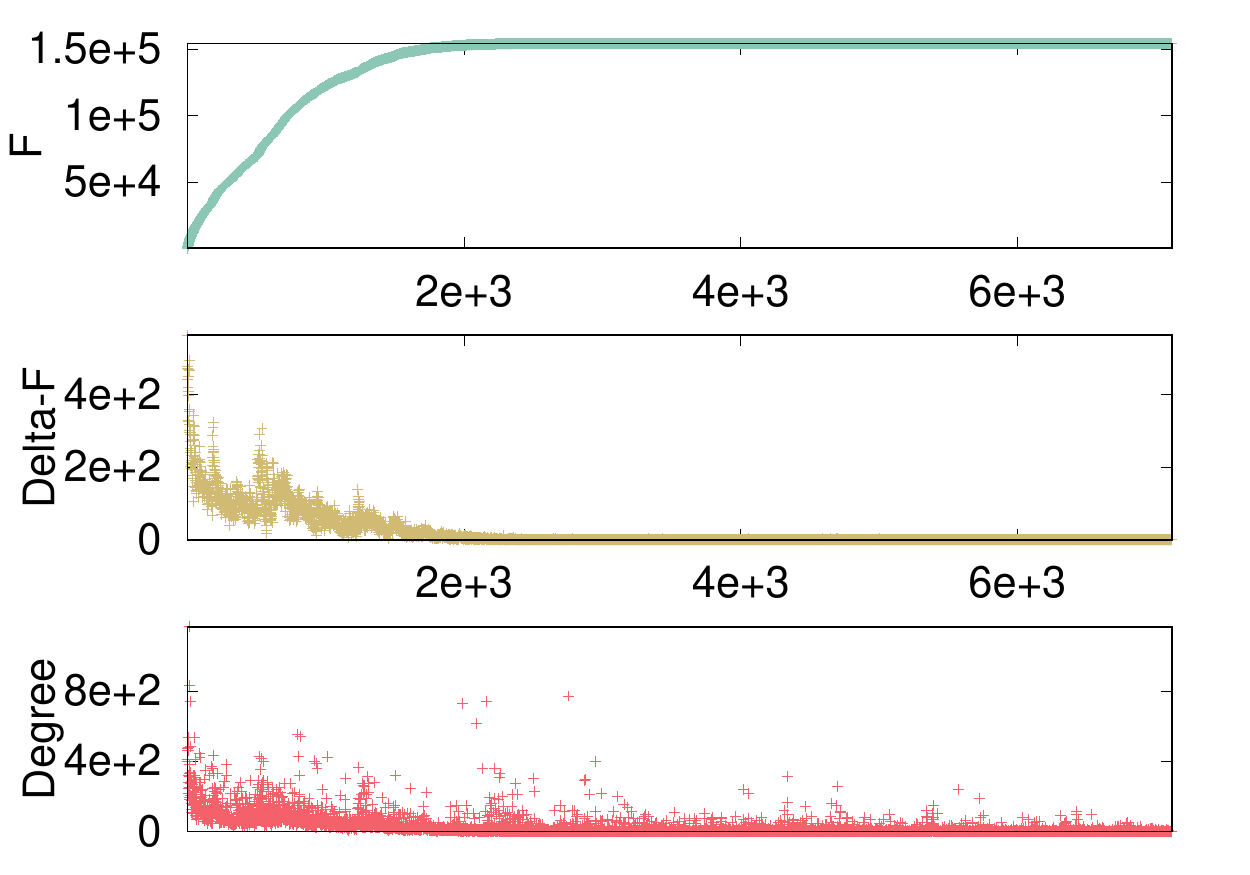} 
  \label{fig:wikivote}}
\subfigure[Air Traffic]{%
  \includegraphics[width=0.475\columnwidth,height=3.1cm]{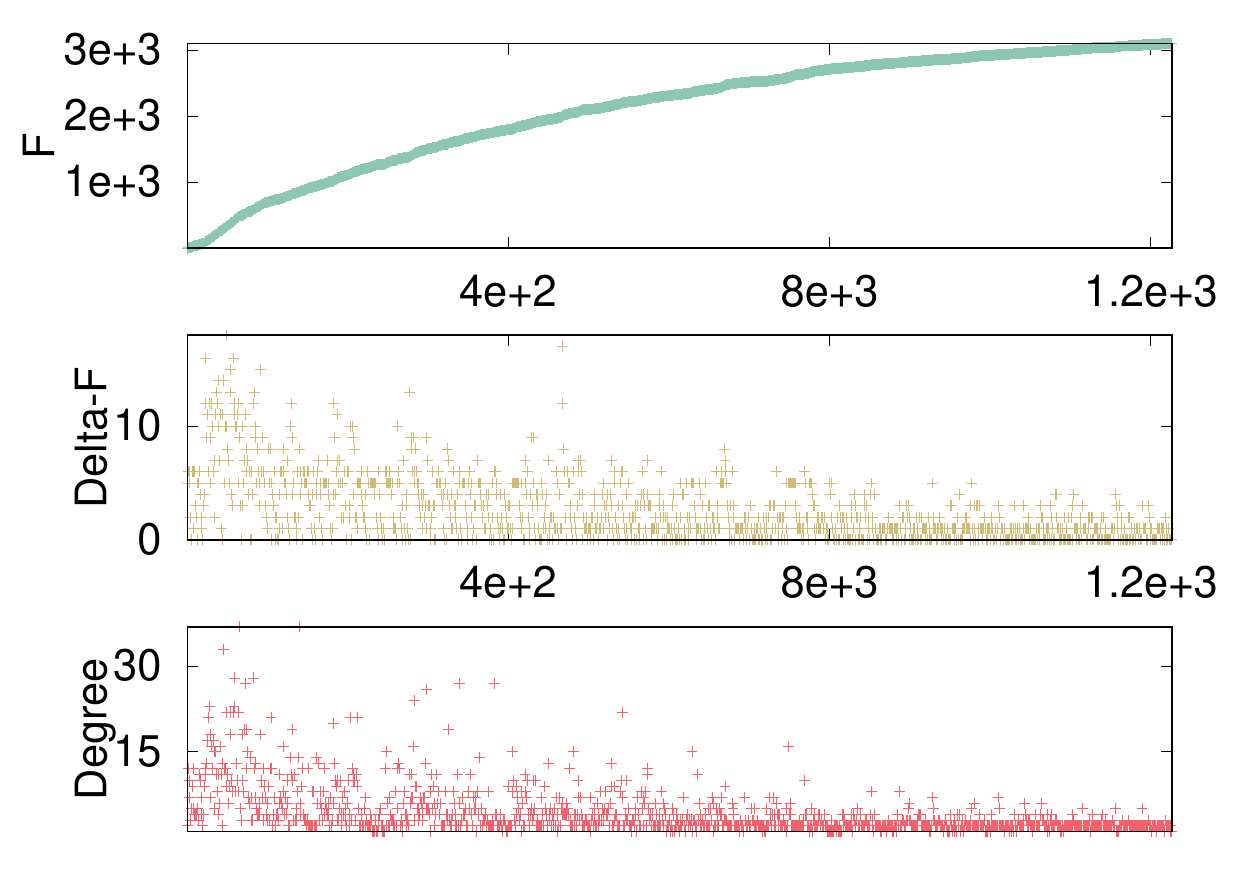}
  \label{fig:air}}
\vspace*{-0.2cm}
\caption{The $F$ score, the increment of $F$ score and the vertex degree of the permutation results of \Gorder on two real graphs: Wiki-Vote and Air Traffic. }
\vspace*{-0.2cm}
\label{fig:adaptTrain}
\end{figure}

\section{RL-based Training}
\label{sec:RL}

In this section, we discuss how we adjust the sampling probability 
adaptively. As shown in Section~\ref{sec:base}, the feasible input of $\QNet$ 
is all possible combinations of the vertices with size $w-1$, which indicates that it is intractable to enumerate all
the partial solutions in training \BaseNet. To alleviate the combinatorial explosion issue, we can employ the sampling strategy in the training space to reduce the candidate partial solutions. 

By intuition, high-degree vertices sharing many common neighbors play
a significant role in constructing high-quality feasible solutions.  However, since the graph ordering is a discrete optimization problem, the significance of a discrete element that contributes to a solution can be very different from others, and there exists skewness regarding such significance among all elements.

To validate this claim, we report the permutation results of two real graphs: Wiki-vote \cite{snapnets} and Air Traffic~\cite{konect} in Fig.~\ref{fig:adaptTrain}. Here, the horizontal axis is the linear graph ordering. In vertical, we show the $F$ score, the increment of $F$, and the degree of the vertex from top to bottom in three subfigures, respectively.
As shown in Fig.~\ref{fig:adaptTrain} we can observe that the patterns of the two solutions are very different.  For Wiki-vote, over $2/3$ vertices in the ordering almost make no contribution to enlarging the $F$ score, while it is not the case for Air Traffic, as vertex adding into the permutation, the $F$ score increases continuously. 
Interestingly, it is counterintuitive that the high-degree vertices do not always make a significant contribution to increasing the $F$ score.

This result inspires us that what if adjusting the sampling probability dynamically during the training phase. This further motives us to use RL framework to tune sampling probability. Therefore, we model the tuning sampling probability in an MDP as follows:


\stitle{States}: a state $s$ is a vector of sampling probability $\prob  \in (0, 1)^n $, $\prob
=\{\prob(v_{i})\}_{i=1}^{n}$  where $n$ is the number of
vertices. $\prob(v_{i})$ is the probability of sampling vertex
$v_i$ for one training instance and ${\bf \prob}$ is normalized, i.e.,
$\Sigma_{i = 1}^{n} \prob(v_{i}) = 1$. It is worth mentioning that the initialization of ${\bf \prob}$ cannot be random. What the Policy Network does is to adjust the experienced-based ${\bf \prob}$ slightly instead of learning it directly without any bias, which is extremely difficult for on-policy RL in an online fashion.

\stitle{Actions}: an action is a vector of 0-1, ${\bf a}= \{
a(i)\}_{i=1}^{n}$ $\in \{0, 1\}^n$. Each element $a(i)$ denotes a tuning
action imposed on $\prob(v_i)$, i.e., increase~(0) or decrease~(1).  There is
a tuning rate $\lambda \in \mathbb{R}$ to control the constant delta
changes of each element $\prob(v_i)$ for a given $a(i)$. 
The adjustment is specified in Eq.~(\ref{eq:prob_tune}).
The final step of an adjustment is the $L1$ normalization of ${\bf \prob}$.

\begin{equation}
{
{\prob(v_i)} =  
\begin{cases} 
 \prob(v_i) + \lambda, & a(i) = 0 \\
 \prob(v_i) - \lambda, & a(i) = 1
\end{cases}
}
\label{eq:prob_tune}
\end{equation}

\stitle{Transition}: given the state-action pair $({\bf \prob_t}, {\bf a_t}
)$  at time $t$, the transition to next state ${\bf \prob_{t+1}}$, is
deterministic.

\stitle{Reward}: the reward $r_t = r({\bf \prob_t}, {\bf a_t} )$ reflects
the benefits of updating a state ${\bf \prob_t}$ by an action ${\bf a_t}$
at time $t$.  We use the evaluation results (e.g., root mean square
error, cross entropy, etc.) of \BaseNet trained on
the training data generated by updating ${\bf \prob_t}$ as $r_t$. To avoid the over-fitting of \BaseNet, instead of sampling the evaluation set $D_{\Dev}$ according to ${\bf \prob_t}$ directly, we adopt the fixed degree sampler, i.e., the initial ${\bf \prob}$ to sample the first vertex and generate the optimal partial solution with window size $w$.  The evaluation is performed on a selected evaluation set, $D_{\Dev}$, after training \BaseNet a certain number of steps. 
The cumulative reward is computed by adding future reward
discounted by a discounting factor $\gamma \in [0, 1]$.

%

\stitle{Policy}: A policy function $\pi({\bf a_t} | { \bf
\prob_t})$ specifies a tuning operation ${\bf a_t}$ to perform on a
given state ${\bf \prob_t}$. Here we employ a parametric function $\pi_{\Theta'}({\bf a_t} | { \bf \prob_t})$ with parameter $\Theta'$ to analog output of the policy. This model can be any any multi-label classification
model~\cite{DBLP:journals/jdwm/TsoumakasK07}, e.g., multilayer 
perceptron~(MLP), which takes ${\bf \prob}$ as input and outputs the
probability of an action ${\bf a}$.  We use $\pi_{\Theta'}({\bf a_t} |
{s_t})$ to denote the output probability at time $t$. To
distinguish from \BaseNet $\QNet$, we use $\Pi(s; \Theta')$ to denote
the Policy Network.


\subsection{Train Tuning Policy}
To find the optimal policy, there are two kinds of RL algorithms~\cite{DBLP:books/lib/SuttonB98}. 
One is value-based, e.g., Q-learning, which learns the optimal state-action values in Eq.~(\ref{eq:bellman}) and the corresponding optimal policy is to take the best action in any state. 
The other is policy-based, which learns the optimal policy directly from a collection of policies. 
Since the policy-based RL algorithm is more suitable for high dimensional action space and has better convergence properties, 
we use a model-free, policy-based RL algorithm to train the Policy
Network directly. 
The objective is to find the optimal policy $\pi^*$ by
maximizing the following expected accumulated and discounted rewards
of Eq~(\ref{eq:loss}), which is estimated by sampling state-action trajectory
$\varsigma$~$(s_0, a_0, r_0, s_1, a_1,$ $...)$ of $T$ steps.
\begin{equation}
{\small
J(\Theta') = \mathbb{E}[ \Sigma_{t=0}^{t=T} \gamma^t r_t | \pi_{\Theta'} ] 
= \mathbb{E}_{\varsigma \sim \pi_{\Theta'}(\varsigma)}[ \Sigma_{t=0}^{t=T} \gamma^t r_t ]
}
\label{eq:loss}
\end{equation}
The gradient of the Eq.~(\ref{eq:loss}) is formulated using the
\REINFORCE algorithm~\cite{DBLP:journals/ml/Williams92} as given in
Eq.~(\ref{eq:gradient}). 
\begin{equation}
{\small
\nabla J(\Theta') = \mathbb{E}_{\varsigma \sim \pi_{\Theta'}(\varsigma)}[ \nabla \log \pi(s_t) (\Sigma_{t=0}^{t=T} \gamma^t r_t - b(s_t)] 
}
\label{eq:gradient}
\end{equation}
Here, 
as the raw reward of a trajectory, $\Sigma_{t=0}^{t=T} \gamma^t r_t$, is always positive, 
a baseline function $b(s)$ is introduced to reduce the variance of the gradients. 
The formula $\Sigma_{t=0}^{t=T} \gamma^t r_t - b(s)$ indicates whether a reward is better or worse than the expected value we should get from state $s$.  
For the baseline function $b(s)$,  we adopt the
widely-used moving average of the historical rewards, i.e., the
evaluation result of \BaseNet.

\begin{algorithm}[t]
{
\caption{ Train the Policy Network $\Pi(s;\Theta')$}
\label{alg:Reinforce}
\begin{algorithmic}[1]
\STATE \textbf{Input}: graph $G$, initial state ${\bf \prob}_{0}$, evaluation set $D_{\Dev}$, learning rate $\alpha$, discounting factor $\gamma$
\STATE Initialize $\Pi(s;\Theta')$ \label{lst:line:init}
\FOR {each RL step } \label{lst:line:start}
\FOR {$t$ = 0 \TO $T$} \label{lst:line:samplestart}
\STATE{Sample $ a_{t} \propto \pi_{\Theta'}({\bf a_t}|s_t)$, where $s_t = {\bf \prob_t}$ }  \label{lst:line:sample}
\STATE{ Compute ${\bf \prob_{t+1}}$ by ${\bf a_t}$ and normalization } \label{lst:line:update}
\STATE{Train and update $\QNet$ based on $D_{t+1}$, where $D_{t+1}$ is sampled from $G$ by ${\bf \prob_{t+1}}$ } \label{lst:line:trainbase}
\STATE{Evaluate $\QNet$, and receive the reward $r_t$ computed on $D_{\Dev}$} \label{lst:line:eval}
\ENDFOR
\FOR {$t$ = 0 \TO $T$}
\STATE {Compute cumulated reward $R_{t} = \Sigma_{i  = t}^{i = T} \gamma^{i-t} r_i$ } \label{lst:line:computereward}
\STATE { $\Theta' \leftarrow \Theta' + \alpha (R_t - b(s_t))\nabla \log \pi(s_t) $} \label{lst:line:gradientascent}
\ENDFOR
\ENDFOR
\STATE \textbf{Output}: the Policy Network  $\Pi(s; \Theta')$
\end{algorithmic}
}
\end{algorithm}

The training algorithm of the Policy Network is given in
Algorithm~\ref{alg:Reinforce}.  It takes a graph $G$, an initial and
experience-based sampling probability ${\bf \prob_0}$, the evaluation set
$D_{\Dev}$, and the learning rate of RL as input, and trains Policy
Network $\Pi(s;\Theta')$ by interacting with the training process of Deep Order
Network $\QNet$.  
The intuition is if the reward of an action is positive, it indicates the action is good and the gradients should be applied to make the action even more likely to be chosen in the future. However, if the reward is negative, it indicates the action is bad and the opposite gradients should be applied to make this action less likely in the future.

We briefly explain the algorithm below.  First,
Policy Network $\Pi(s;\Theta')$ is initialized~(line~\ref{lst:line:init}). The
training of policy gradient starts at
line~\ref{lst:line:start}. Before that, the training of \BaseNet
has started several steps to collect enough evaluation results for
computing the baseline.  In each RL step, the algorithm updates $\Pi(s;\Theta')$
once by one Monte Carlo sampling of a trajectory of length $T$
(line~\ref{lst:line:samplestart}).  For each time $t$, it repeats the
following 4 steps one by one.  First, draw a random action ${\bf a_t}$
based on the probability output by Policy Network, i.e.,
$\pi_{\Theta'}({\bf a_t} |{\bf \prob_t})$~(line~\ref{lst:line:sample}). Instead of
directly choosing the action with the highest probability, the random
action manages a balance between exploring new actions and exploiting
the actions which are learned to work well.  Second, apply the action
${\bf a_t}$ on the state ${\bf \prob_t}$ by imposing an
increment/decrement of $\lambda$~(line~\ref{lst:line:update}), to
generate the next state ${\bf \prob_t}$.  Third, generate the
training data set $D_{t+1}$ of \BaseNet by sampling $G$ with probability ${\bf
  \prob_{t+1}}$, and feed $D_{t+1}$ to train $\QNet$~(line~\ref{lst:line:trainbase}).  Fourth, after training
$\QNet$ in a certain number of steps, we use the evaluation set
$D_{\Dev}$ to evaluate \BaseNet, and collect the evaluation
result as reward $r_t$~(line~\ref{lst:line:eval}).  When a trajectory
is simulated, we compute the cumulative
rewards~(line~\ref{lst:line:computereward}) and apply gradient ascent
to update the Policy Network $\Pi(s;\Theta')$~(line~\ref{lst:line:gradientascent}).


\section{Experimental Studies}
\label{sec:exp}
In this section, we present our experimental evaluations. 
First, we give the specific setting of the testing including datasets, settings of the models and training. Then, we compare the proposed model: \BaseNetPlus with the state-of-the-art algorithmic heuristic, conduct an A/B testing to validate the effect of Policy Network, and observe the performance of models as $w$ varies. Finally, two case studies of compressing graph and deriving a competitive edge partitioning from model-based ordering are presented. 

\subsection{Experimental Setup}

\begin{table}[t]
{\footnotesize 
\caption{Datasets}
\label{tbl:datasets}
\vspace*{-0.3cm}
\begin{center}
    \begin{tabular}{|l|r|r|r|r|} \hline
    {\bf Graphs}              & {\bf $|V|$ } & $ {\bf |E|} $   & {\bf Density}  & {\bf Description}
    \\ \hline\hline                                               
   Wiki-vote~\cite{snapnets}  & 7,115  &  103,689  & $0.0020$  &vote graph \\ \hline 
   Facebook~\cite{snapnets}   & 4,039  &  88,234    & $0.0108$ & social graph\\ \hline 
    p2p~\cite{snapnets}       &  6,301 &  20,777    & $0.0005$ & file sharing graph\\\hline 
    Arxiv-HEP~\cite{snapnets} & 9,877  &  25,998   & $0.0003$ & co-authorship graph \\ \hline
    Cora~\cite{konect}        & 23,166 &  91,500   & $0.0001$ & citation graph\\\hline 
    PPI~\cite{snapnets}       & 21,557 &  342,353  & $0.0015$ &  biological graph\\\hline 
    \hline                                                     
    PL10K\_1.6                & 10,000 & 121,922   & $0.0024$ &  \multirow{3}{*}{power-law graph}\\ \cline{1 - 4}
    PL10K\_1.8                & 10,000 & 58,934    & $0.0011$ &     \\ \cline{1 - 4}
    PL10K\_2.0                & 10,000 & 31,894    & $0.0006$ &      \\\cline{1 - 4}
    \hline \hline                                                          
    ER10K\_0.02               & 10,000 & 1,000,337 & $0.0200$ & \multirow{3}{*}{ ER graph}  \\ \cline{1 - 4}
    ER10K\_0.05               & 10,000 & 2,498,836 & $0.0500$ &  \\ \cline{1 - 4}
    ER10K\_0.1                & 10,000 & 5,004,331 & $0.1000$ &  \\ \hline

   \end{tabular}
\end{center}
}
\vspace*{-0.2cm}
\end{table}

\stitle{Datasets:} 
We use six real graphs collected from Stanford SNAP~\cite{snapnets} and KONECT (The Koblenz Network Collection)~\cite{konect}: \emph{Wiki-vote} is the Wikipedia adminship vote network. 
\emph{Facebook} is the Facebook friendship network. 
\emph{p2p} is a snapshot of the Gnutella peer-to-peer file sharing network. 
\emph{Arxiv-HEP} is the Arxiv collaboration network between authors in the field of High Energy Physics. 
\emph{Cora} is the cora scientific paper citation network. 
\emph{PPI} is a protein-protein interaction network where vertices represent human proteins and edges represents physical interactions between proteins.

In addition, to validate the performance of our models on graphs with different structures explicitly,  
we generate six synthetic graphs by SNAP random graph generator: power-law and \ERGraph. 
For power-law, we choose three power-law distributions with shape parameter $\gamma \in $ \{1.6, 1.8, 2.0\} and adopt Newman's method \cite{newman2001random} to generate three power-law graphs: PL10K\_1.6, PL10K\_1.8 and PL10K\_2.0. 
For \ERGraph, we utilize the parameter $p \in  \{0.02, 0.05. 0.1\}$ which indicates the probability of edge existence between two vertices to generate three synthetic graphs with different densities, that are, ER10K\_0.02,ER10K\_0.05 and ER10K\_0.1. All synthetic graphs have 10,000 vertices. 
Table~\ref{tbl:datasets} summarizes the information of these real and synthetic graphs. 

\begin{figure}[t]
\centering
\begin{subfigure}[Order Equivalence]{
\label{fig:vertexreduction:1} {\scriptsize
    \begin{tikzpicture}[
            > = stealth, 
            shorten > = 1pt, 
            auto,
            node distance = 1.2cm, 
            semithick 
        ]

        \tikzstyle{every state}=[
            draw = black,
            thick,
            fill = white,
            minimum size = 4mm
        ]

        \node[state] (s) {$u$};
        \node[state] (v1) [above right of=s, fill = yellow] {$v_1$};
        \node[state] (v2) [right of=s, fill = yellow] {$v_2$};
        \node[state] (v3) [below right of=s, fill = yellow] {$v_3$};
        \node[state] (u1) [above left of=s] {$$};
        \node[state] (u2) [below left of=s] {$$}; 

        \path[->] (s) edge node {} (v1);
        \path[->] (s) edge node {} (v2);
        \path[->] (s) edge node {} (v3);
        \path[->] (u1) edge node {} (s);
        \path[->] (u2) edge node {} (s);
        \path[->] (u1) edge node {} (u2); 
        \path[->] (u2) edge node {} (u1); 
    \end{tikzpicture}
    }}
\end{subfigure}
\begin{subfigure}[Merged Vertex]{
\label{fig:vertexreduction:2} {\scriptsize
    \begin{tikzpicture}
    	[
            > = stealth, 
            shorten > = 1pt, 
            auto,
            node distance = 1.3cm, 
            semithick 
        ]

        \tikzstyle{every state}=[
            draw = black,
            thick,
            fill = white,
            minimum size = 4mm
        ]

        \node[state] (s) {$u$};
        \node[state] (v) [right of=s, fill = yellow] {$v^*$};
        \node[state] (u1) [above left of=s] {$$};
        \node[state] (u2) [below left of=s] {$$}; 

        \path[->] (s) edge node {} (v);
        \path[->] (u1) edge node {} (s);
        \path[->] (u2) edge node {} (s);
        \path[->] (u1) edge node {} (u2); 
        \path[->] (u2) edge node {} (u1);
    \end{tikzpicture}
    }}
\end{subfigure}
\caption{Reducing the Sampling Classes $V(G)$} 
\label{fig:vertexreduction}
\vspace*{-0.3cm}
\end{figure}
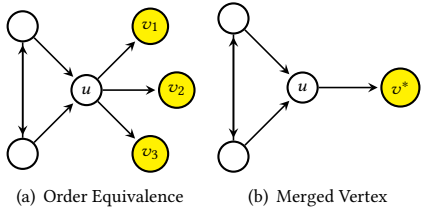

\stitle{Reduce the Sampling Classes:}
Recall that \BaseNet takes a vertices set sampled from $V(G)$ as input to predict the likelihood of appending each vertex to the permutation. 
Based on the RL-based sampling probability tuning approach we have explored,  we can further reduce the sampling classes $V(G)$ by a simple but effective strategy. 
The reduction is based on the automorphism of vertices on the trig of the graph. We use an example in Fig.~\ref{fig:vertexreduction} to illustrate this reduction. 
In Fig.~\ref{fig:vertexreduction:1}, vertices $v_1$, $v_2$ and $v_3$ have a common in-neighbor $u$, which is also their only neighbor. 
For $v_i ~(i = 1, 2, 3)$, we have $\SimFunc(v_i, u) = 1$ and the $\SimFunc$ value is $0$ for all the other vertices. 
$v_1$, $v_2$ and $v_3$ are automorphism so that permuting them can be regarded as permuting one vertex $v^*$ three times, where $\SimFunc(v^*, u) = 1$ and $\SimFunc$ is 0 for the other vertices.  
In other words, if three positions for the three vertices $v_i$ are targeted in the permutation, the $F$ value will be equivalent for all the $3!$ assignments of $v_i$. 
In this vein, we merge $v_1$, $v_2$ and $v_3$ to one vertex $v^{*}$ as shown in Fig.~\ref{fig:vertexreduction:2}. Sampling vertex sets and inference to generate permutation can be directly conducted on the merged graph (Fig.~\ref{fig:vertexreduction:2}). 
Each time the model selects $v^{*}$ to extend the permutation, we insert a vertex randomly drawn from $v_1$, $v_2$, $v_3$. For real-world graph, due to its power-low degree distribution, our experiments show this preprocessing can reduce about 9\% vertices on average.

\stitle{Model settings:}
In \BaseNet, both the $\phi$ and $\rho$ networks are two-layer MLP. We use the \kw{SUM} as the pooling layer to add up the representations of $\phi$ and the output is further processed by $\rho$.
The input of $\phi$ is an identity vector representation, $\{0, 1\}^n$, of a partial solution with size $w-1$ and the output is a probability distribution $p({\bf v}|S) \in (0,1)^n$ generated by a softmax function. 
All the hidden units are activated by ReLU function $ReLU(x) = \max(0, x)$. 

The Policy Network is a multi-label classifier of two-layers MLP. 
The input is the state ${\bf \prob_{t}} \in (0, 1)^n $ and the output layer predicts the probability to perform action ${\bf a_t}$, i.e., a vector $(0, 1)^n$ by sigmoid activation. 
Similar to \BaseNet, the hidden units are activated by ReLU function.

\begin{table}[t]
{\footnotesize 
\caption{The Hyper-parameter Configuration}
\label{tbl:hyperpara}
\vspace*{-0.3cm}
\begin{center}
    \begin{tabular}{|c|l|c|} \hline
    \multicolumn{2}{|c|}{\bf Hyper-parameters}  & {\bf Values} 
     \\\hline\hline
  \multirow{3}{*}{\BaseNet} & learning rate & $10^{-3} \sim 10^{-4}$\\ \cline{2-3} 
  							   & mini-batch size   & 64 $\sim$ 512  \\ \cline{2-3} 
							    & \# hidden units   & 32, 64, 128, 256  \\ \cline{2-3} 
    						   & global steps & $5 \times 10^{3} \sim \times 10^{6}$ \\\hline
    \hline
   \multirow{5}{*}{Policy Network} &  learning rate  & $10^{-3} \sim 10^{-5}$ \\ \cline{2-3}
   								   & trajectory length $T$  & 1 $\sim$ 10 \\ \cline{2-3}
                                   & RL steps  & 50 $\sim$ 300 \\ \cline{2-3}
                                   & discounting factor $\gamma$  & 0.9, 0.95 \\ \cline{2-3}
                                   & tuning rate $\lambda$  & $ 0.1n \sim 0.2n$ \\ \cline{2-3}
                                   & \# hidden units   & 32, 64, 128, 256  \\ \cline{2-3}

   & \# evaluation set & 2000, 5000  \\ \hline
   \end{tabular}
\end{center}
}
\vspace*{-0.2cm}
\end{table}

\stitle{Implementation and training:} The learning framework is built on Tensorflow~\cite{DBLP:conf/osdi/AbadiBCCDDDGIIK16} 1.8 with Python 2.7. We use Adam~\cite{DBLP:journals/corr/KingmaB14} and RMSProp optimizer to train \BaseNet and the Policy Network, respectively, which are trained interactively as shown in Algorithm~\ref{alg:Reinforce}. 

Table~\ref{tbl:hyperpara} shows the hyper-parameters settings in the training. The models are learned with these parameters tuned in the corresponding experienced range. 
In Table~\ref{tbl:hyperpara}, the global steps and RL steps are the numbers of parameter updating of \BaseNet and the Policy Network, respectively. 
For one RL step, a trajectory of length $T$ \BaseNet training is conducted. To this end, (global steps) / (RL steps $\times T$) \BaseNet steps are trained in one timestamp. 
%
For computing the rewards, we use the Root Mean Square Error (RMSE) as the evaluation metric of the evaluation set. 
The reward is defined as the opposite number of RMSE.
We use the best neighbor heuristic to generate the evaluation set by randomly choose the first vertex with initial $\prob$, the degree distribution.
A bare \BaseNet is trained by setting sampling probability as ${\bf \prob_{0}}$. In this section, we use \BaseNet and \BaseNetPlus to denote the bare \BaseNet model and \BaseNet with Policy Network respectively. 

\begin{table}[t]
{\small
\caption{Result of Graph Ordering}
\label{tbl:graphorder}
\vspace*{-0.3cm}
\begin{center}
    \begin{tabular}{|l|l|r|r|r|} \hline
    {\bf $F(\Phi)$}  & $|V'|$ & \Gorder  & \BaseNet      & \BaseNetPlus
    \\ \hline\hline                                            
   Wiki-vote         & 5,880  & 145,736  & 156,871       &  {\bf 157,669} \\ \hline
   Facebook          & 3,974  & 230,031  & 207,511       & {\bf 234,151}  \\ \hline
   p2p               & 5,624  & 20,472   & 21,422        & {\bf 22,086} \\\hline
   Arxiv-HEP		 & 9,877  & 85,958   & 90,629		 & {\bf 91,001} \\ \hline 	
   Cora              & 22,317 & 98,334   & {\bf 101,063} & 100,966 \\ \hline
   PPI               & 19,041 &  383,343 & 347,237       & {\bf 404,728} \\ \hline
   \hline                                                        
    PL10K\_1.6       & 8,882  & 166,540  & 190,021       & \bf{197,622} \\ \hline
    PL10K\_1.8       & 8,292  & 66,272   & 86,840        & \bf{89,930} \\ \hline
    PL10K\_2.0       & 8,484  & 29,373   & 37,332        & \bf{38,071} \\\hline
    \hline                                                                
    ER10K\_0.02      & 10,000 & 136,925  & 145,084       & {\bf 162,462}  \\ \hline
    ER10K\_0.05      & 10,000 & 615,706  & 673,357       & {\bf 724,743} \\ \hline
    ER10K\_0.1       & 10,000 & 2,319,250& 2,554,032     & {\bf 2,647,686} \\\hline
    \end{tabular}
\end{center}
}
\vspace*{-0.3cm}
\end{table}

\begin{figure*}[t]
\centering
\subfigure[PPI]{%
 \includegraphics[width=0.48\columnwidth]{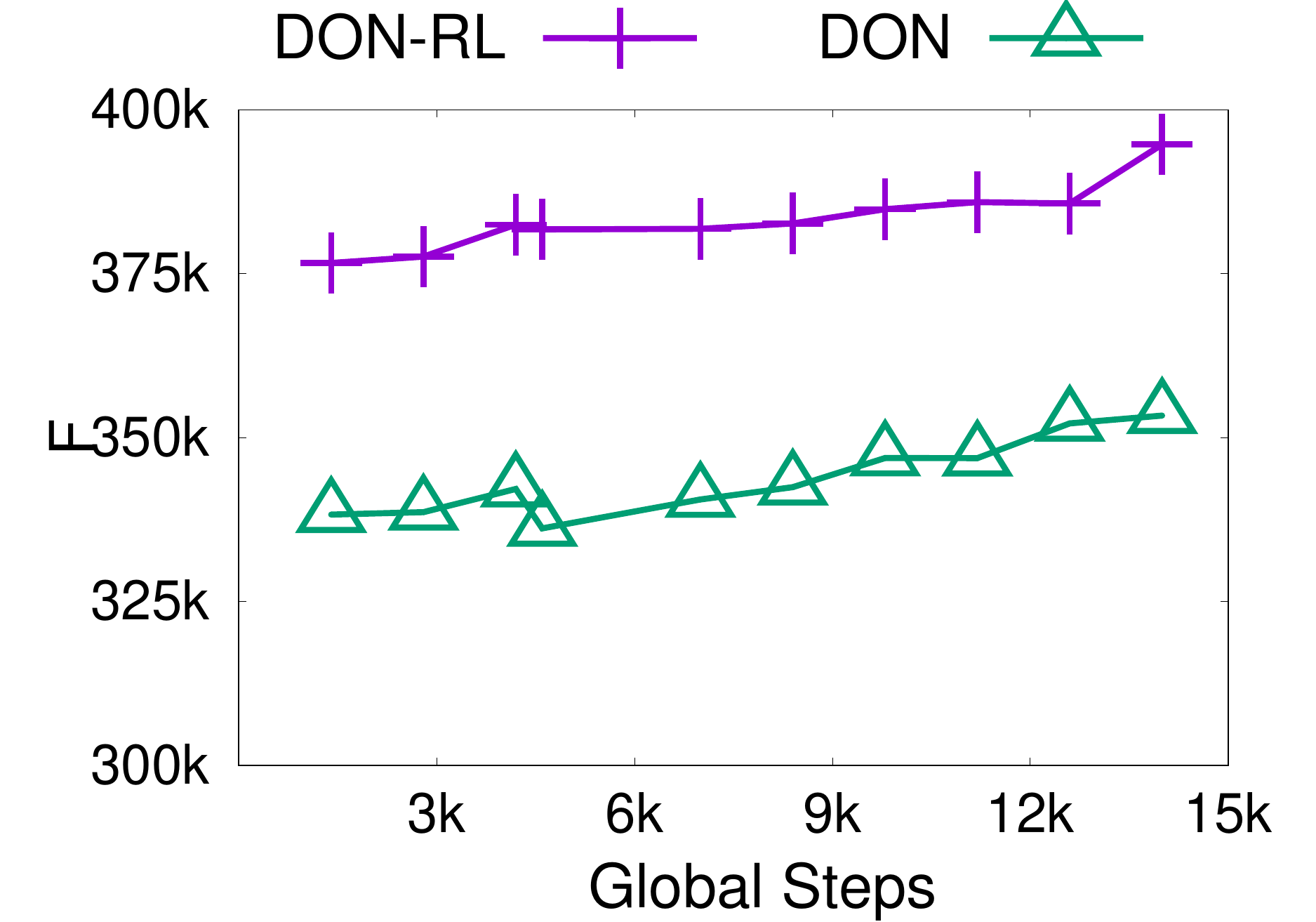}
  \vspace*{-0.2cm}  
  \label{fig:rl:ppi}}
\subfigure[p2p]{%
  \includegraphics[width=0.48\columnwidth]{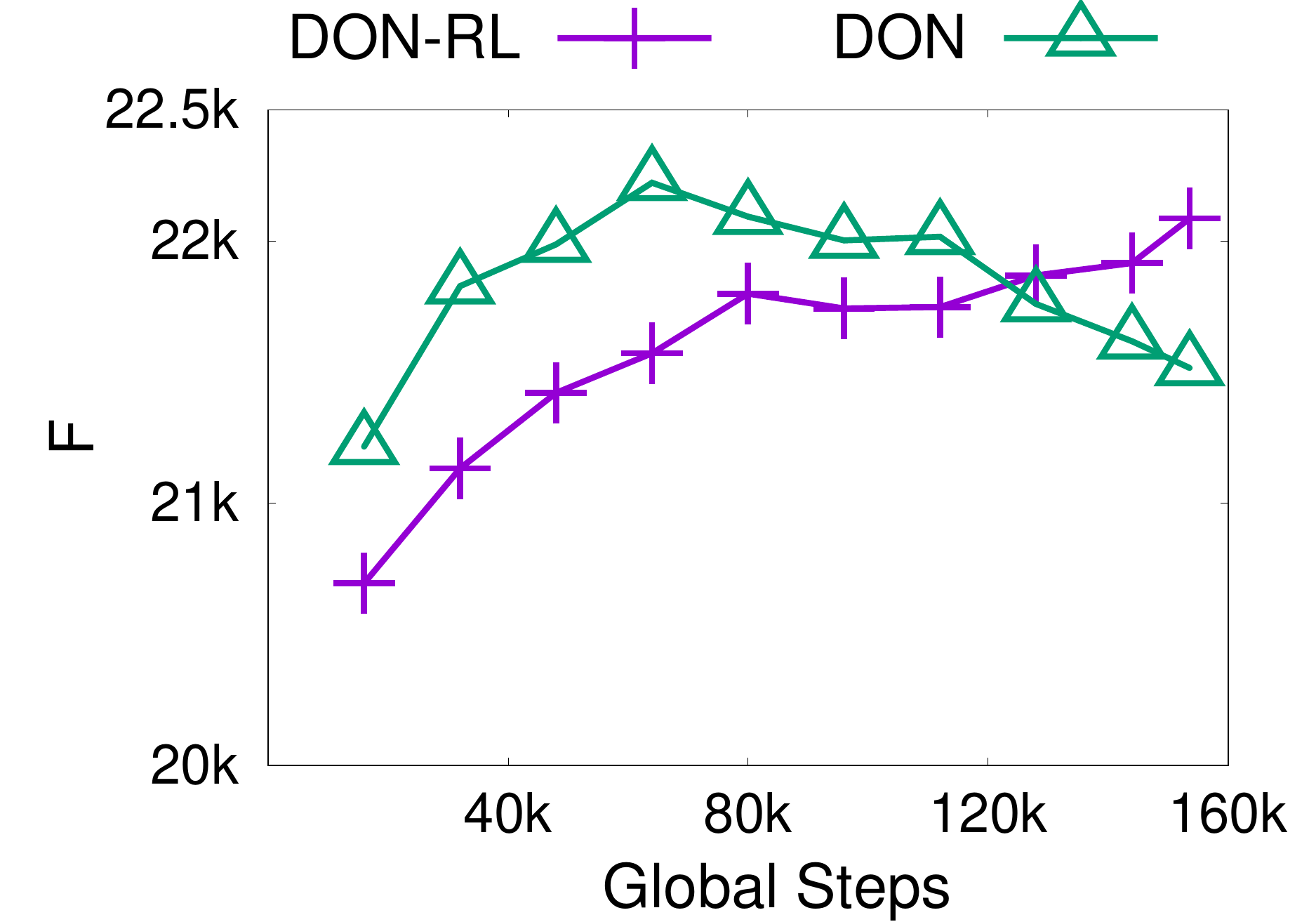}
   \vspace*{-0.2cm}
  \label{fig:rl:p2p}}
\subfigure[Facebook]{%
 \includegraphics[width=0.48\columnwidth]{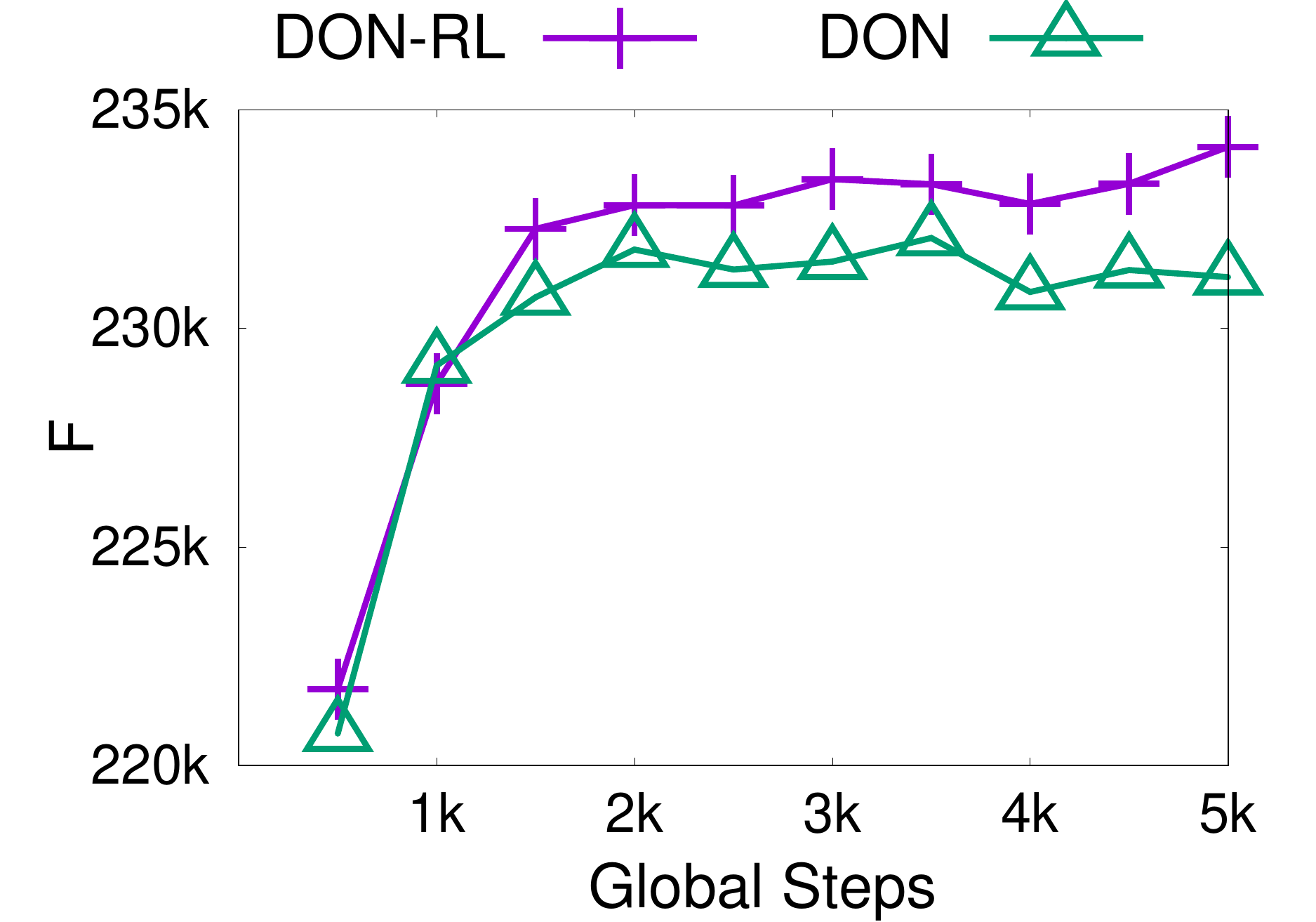} 
  \vspace*{-0.2cm} 
  \label{fig:rl:fb}}
\subfigure[Cora]{%
  \includegraphics[width=0.48\columnwidth]{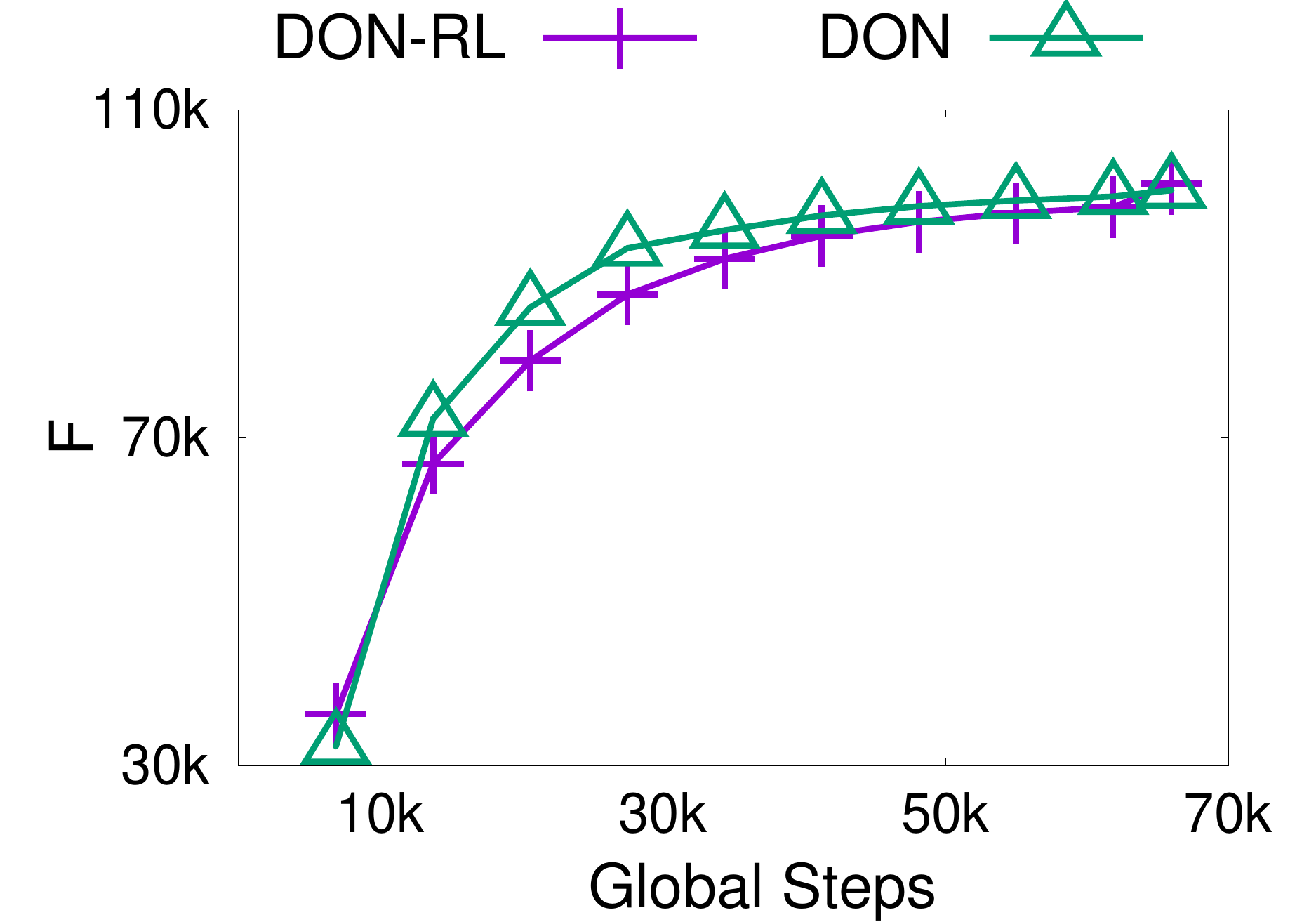}
   \vspace*{-0.2cm} 
  \label{fig:rl:cora}}
\subfigure[Wikivote]{%
  \includegraphics[width=0.48\columnwidth]{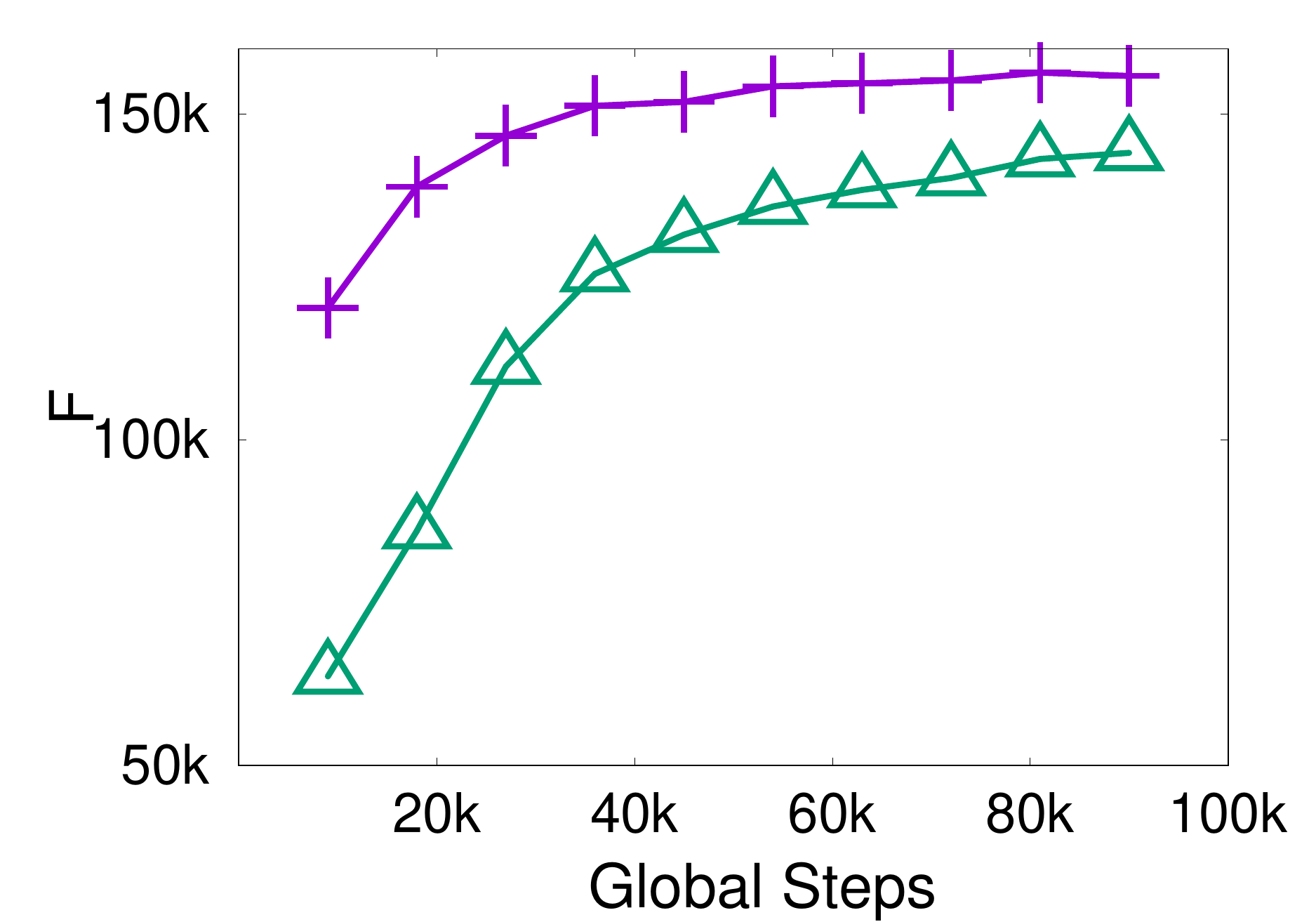}
   \vspace*{-0.2cm} 
  \label{fig:rl:wv}}
\subfigure[PL10K\_1.6]{%
 \includegraphics[width=0.48\columnwidth]{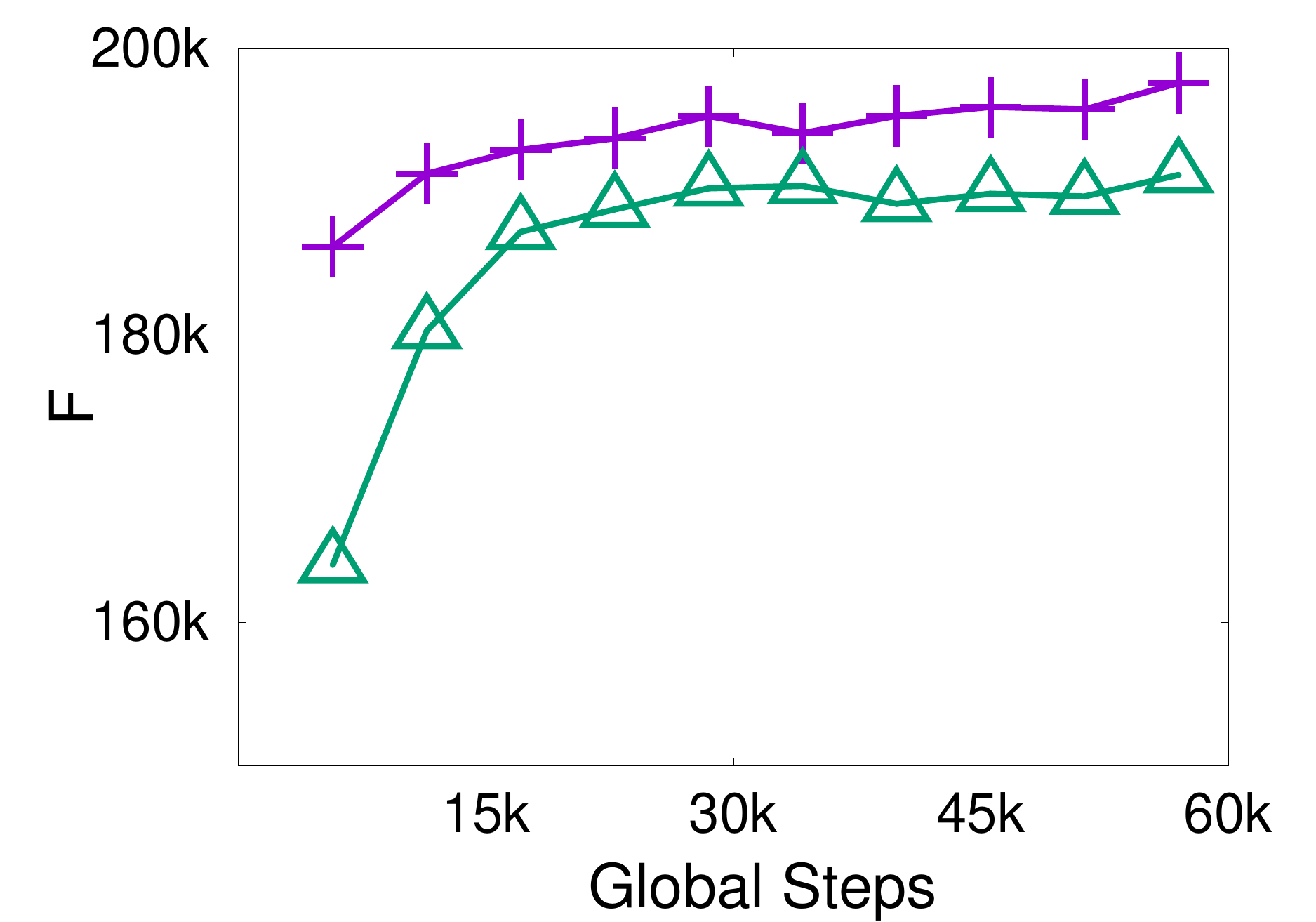} 
 \vspace*{-0.2cm}
  \label{fig:rl:pl10k1.6}}
 \subfigure[PL10K\_1.8]{%
 \includegraphics[width=0.48\columnwidth]{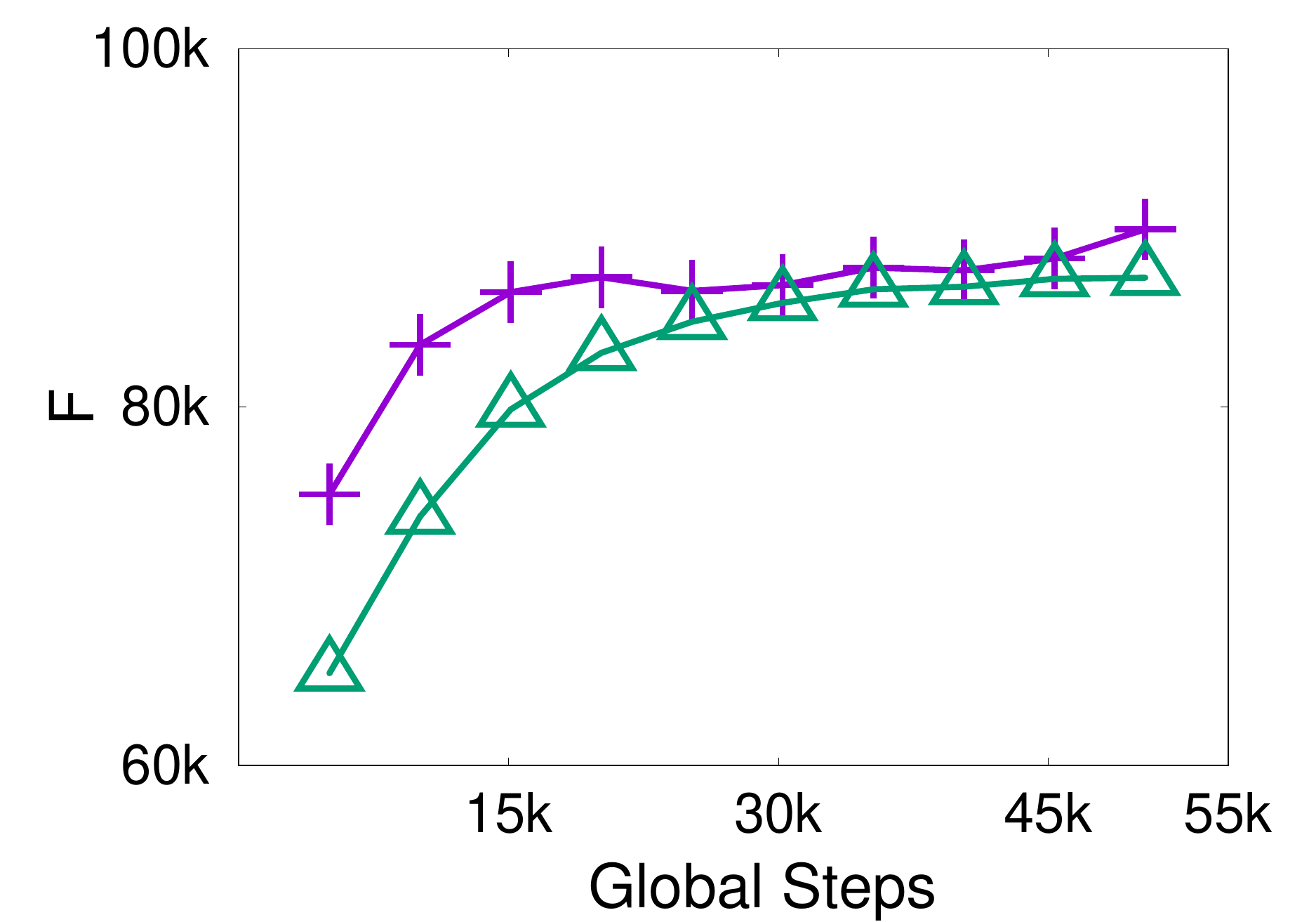} 
 \vspace*{-0.2cm}
  \label{fig:rl:pl10k1.8}}
  \subfigure[PL10K\_2.0]{%
 \includegraphics[width=0.48\columnwidth]{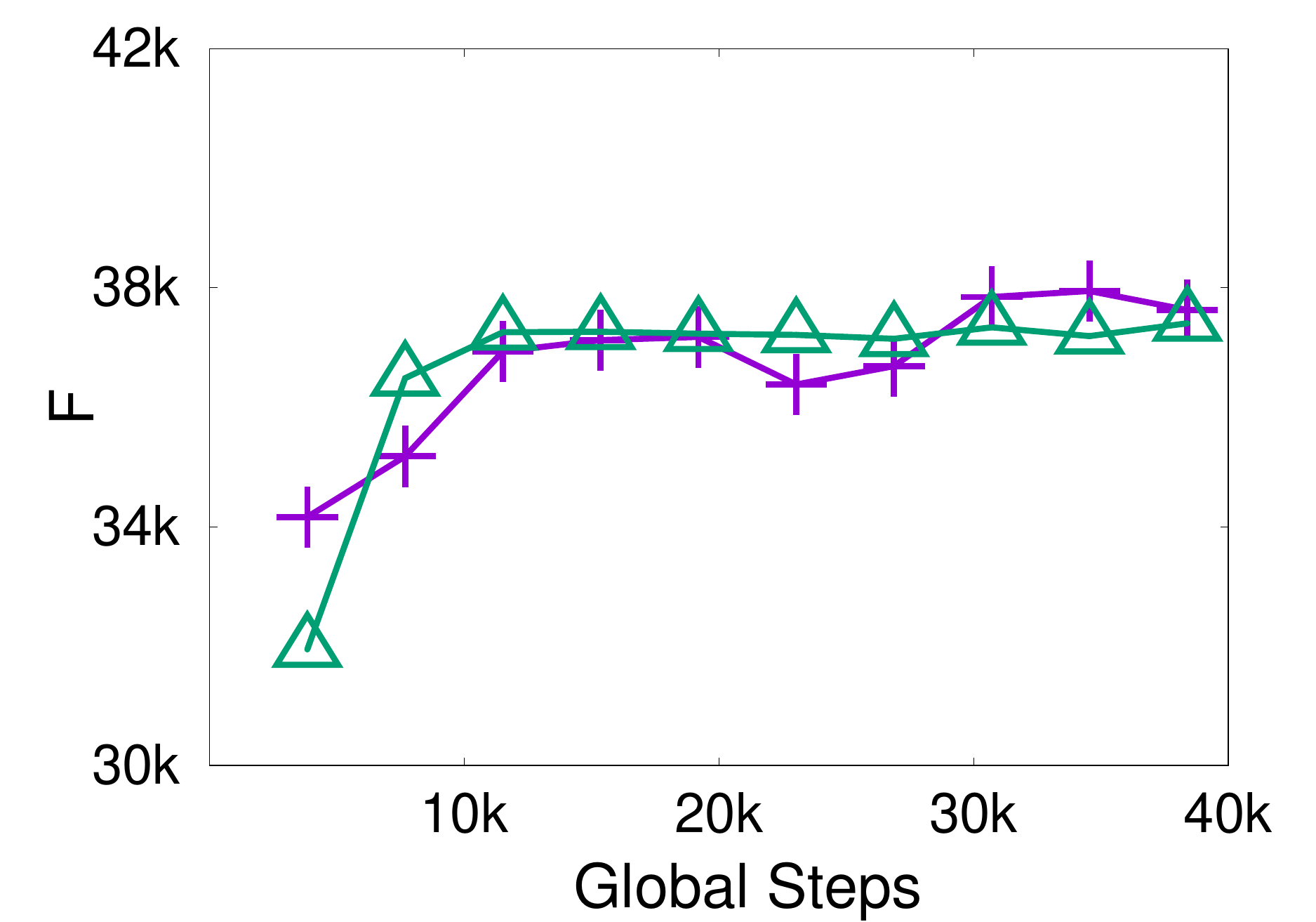} 
 \vspace*{-0.2cm}
  \label{fig:rl:pl10k2.0}}
\subfigure[Arxiv-HEP]{%
  \includegraphics[width=0.48\columnwidth]{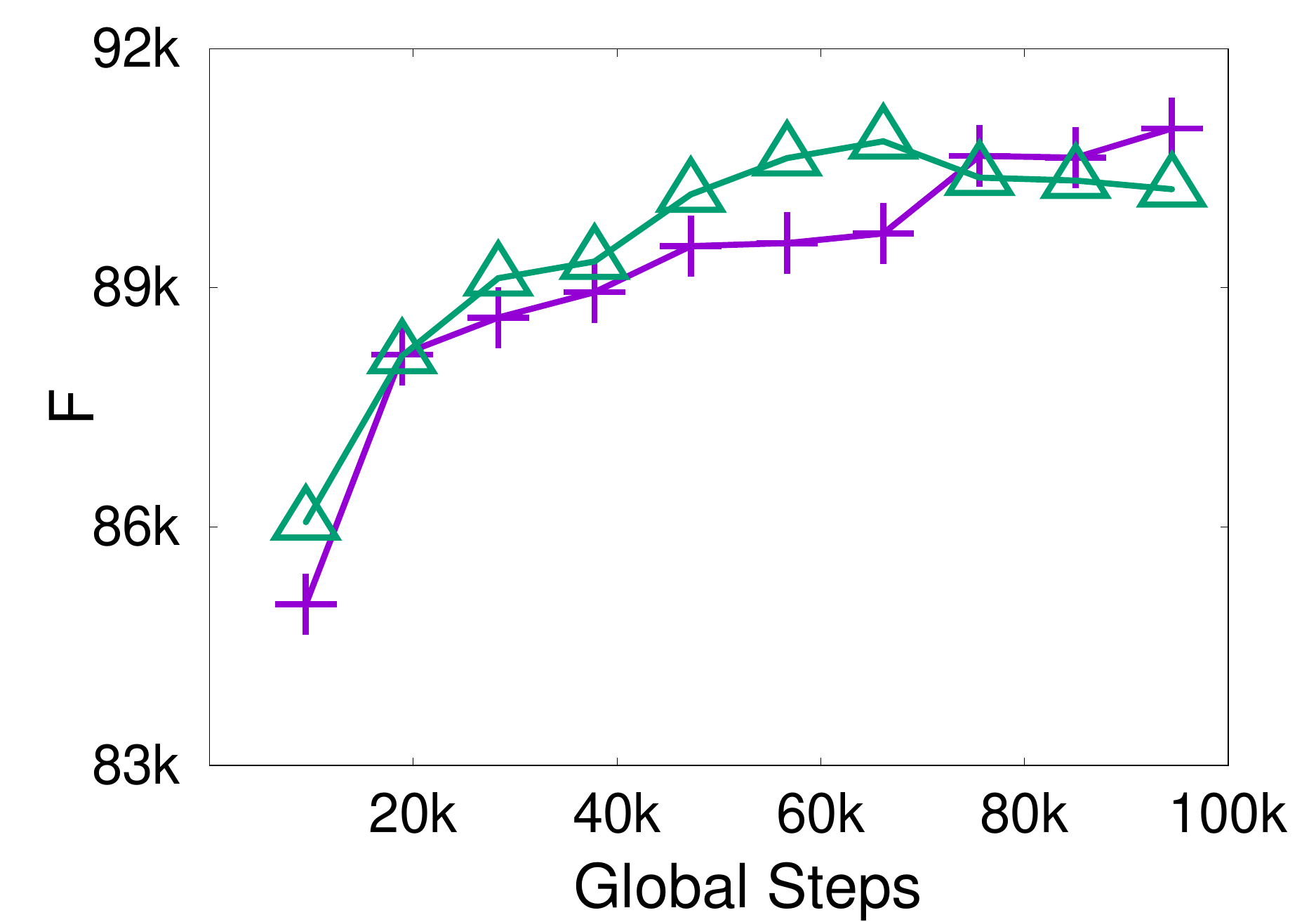}
   \vspace*{-0.2cm} 
  \label{fig:rl:ht}}
\subfigure[ER10K\_0.02]{%
  \includegraphics[width=0.48\columnwidth,height=2.9cm]{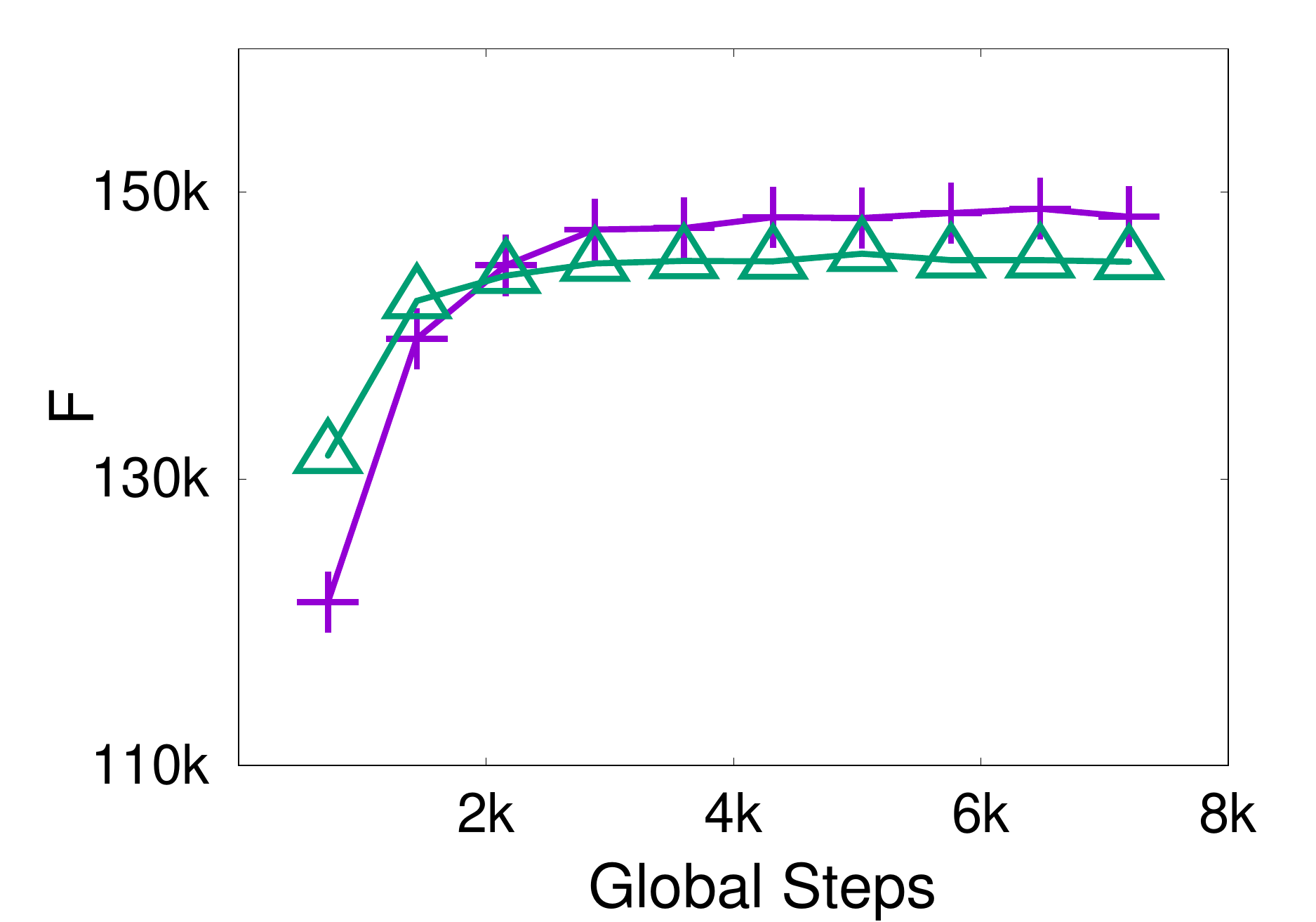}
  \vspace*{-0.2cm}
  \label{fig:rl:er10k0.02}}  
\subfigure[ER10K\_0.05]{%
  \includegraphics[width=0.48\columnwidth,height=2.9cm]{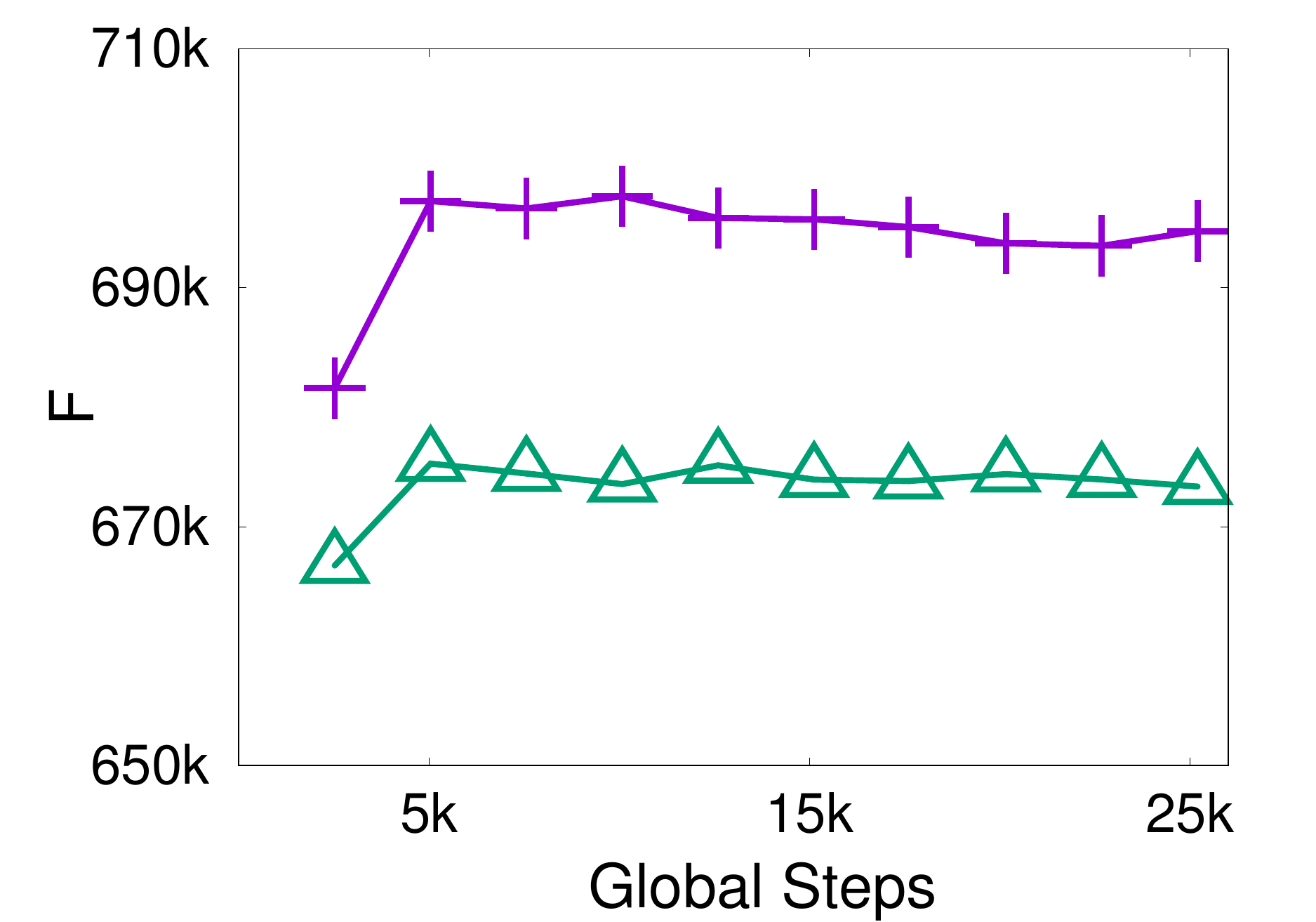}
  \vspace*{-0.2cm}
  \label{fig:rl:er10k0.05}} 
\subfigure[ER10K\_0.1]{%
  \includegraphics[width=0.48\columnwidth,height=2.9cm]{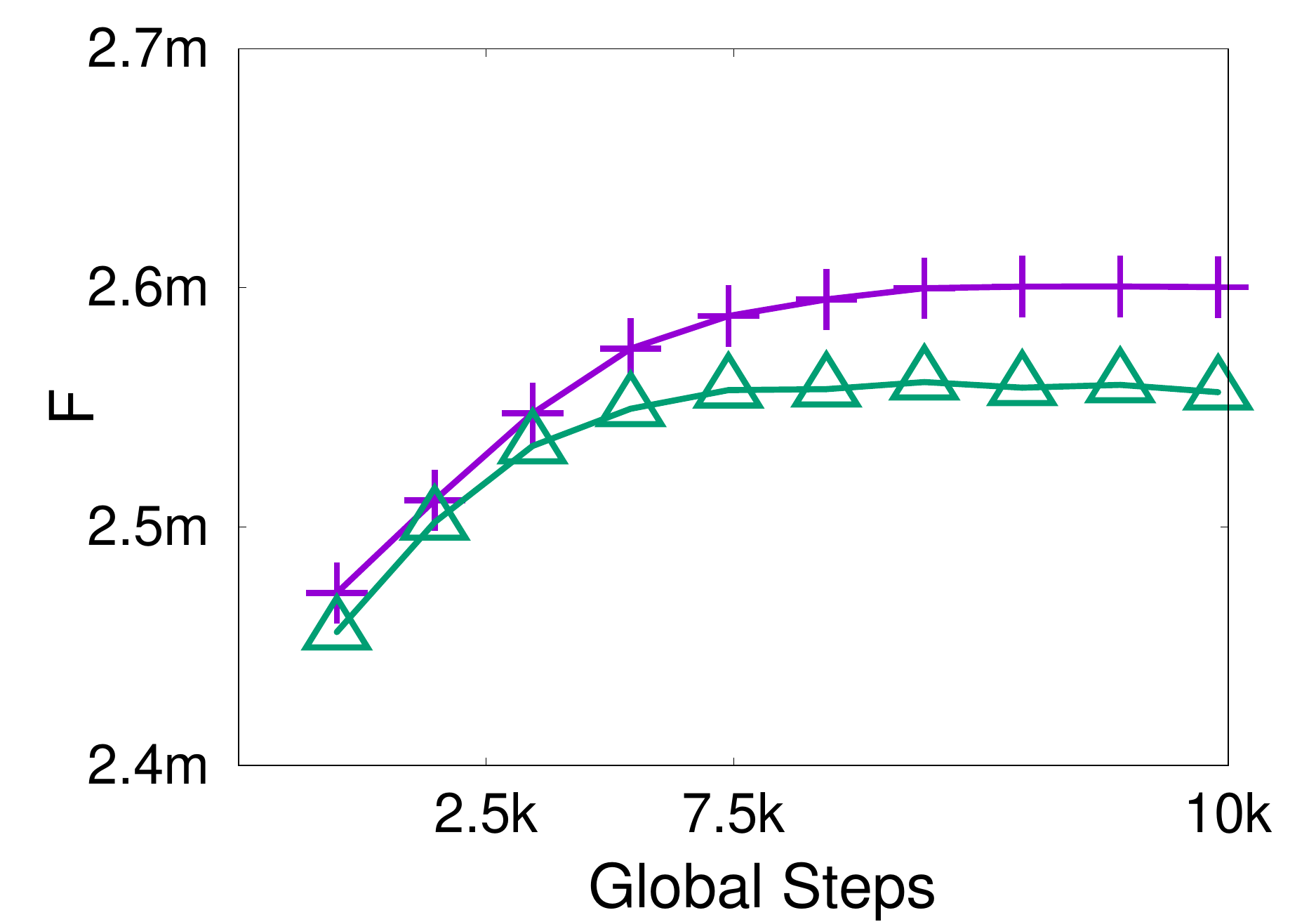}
  \vspace*{-0.2cm}
  \label{fig:rl:er10k0.1}} 
\vspace*{-0.2cm}
\caption{\BaseNet vs. \BaseNetPlus}
\vspace*{-0.2cm}
\label{fig:rl}
\end{figure*}

\subsection{Result of Graph Ordering}
Table~\ref{tbl:graphorder} shows the overall results of the $F$ score of the permutations generated by \BaseNet, compared with the baseline graph ordering algorithm \Gorder~\cite{DBLP:conf/sigmod/WeiYLL16}. 
\eat{
The window size $w$ is fixed to 5, which is commonly used in \cite{DBLP:conf/sigmod/WeiYLL16}. 
We use a simple trick to preprocess the graph to achieve a more compacted model. 
Vertices on the trig of the graph, i.e., vertex whose degree is 1 are merged to one equivalent vertex if they have the same neighbor. 
}
As a preprocessing step, vertices on the trig of the graph, i.e., vertex whose degree is 1 are merged to one equivalent vertex as shown in Fig.~\ref{fig:vertexreduction}.
In Table~\ref{tbl:graphorder}, the column $|V'|$ is the compacted vertex number of the graph. 
We can see that this trick compresses up to 17\% vertices for real and power-low graphs but it is invalid for three \ERGraph graphs due to their degree distribution. 
With regards to maximize $F$, the solutions of \BaseNet and \BaseNetPlus surpass that of the greedy algorithm significantly. 
For the real graphs, all the solutions of \BaseNetPlus outperform that of \Gorder 
from 1.8\% up to 8.2\%.  And the solution of \BaseNet, on the 3 reals graphs, Wiki-Vote, p2p and Cora, the improvements of $F$ are 7.6\%, 4.7\% and 2.8\%, respectively. 
For the synthetic graphs, both the \BaseNet and \BaseNetPlus can generate solutions of much higher $F$ score than \Gorder.  
The \BaseNetPlus performs best and its solutions surpass that of \Gorder up to 35.0\% for powerlaw graph, and 18.6\% for ER graph. 
As the matrix visualization results
shown in Fig.~\ref{fig:matrixvisual}, the model-based approach \BaseNetPlus focuses on the overall performance and can optimize the permutation of the sub-significant vertices. The greedy heuristic is powerful for ordering the top significant vertices while for the sub-significant vertices, it fails to capture the global information as ordering them w.r.t. the common neighbor relations. 
It is easy for \Gorder to stuck in forming local dense areas while incurring sparsity in the global scope.
This phenomenon agrees with our intuition, that is, the greedy algorithm could neglect better solutions in the future scope.

\subsection{Adaptive Training by RL}
In order to validate the effectiveness of the Policy Network, we perform A/B test to observe the effect of RL-based adaptive training, i.e, using the same  hyper-parameters to train a \BaseNet model and a conjuncted \BaseNetPlus model. 
During the overall training process, we generate a solution after per 10\%  global training steps of \BaseNet ends. We directly observe the changes of objective function $F(\Phi)$ as the loss function of the model is not a direct reflection of the objection of this NP-hard  problem. 
Fig.~\ref{fig:rl} presents the results over the 12 graphs in Table~\ref{tbl:datasets}.

We observe that the performance of \BaseNet and \BaseNetPlus are different among different graphs during training. 
In general, imposing RL on the sampling probability adjustment improves the performance of  \BaseNet in most cases. 
As shown in Fig.~\ref{fig:rl:ppi}, \ref{fig:rl:fb}, \ref{fig:rl:wv}-\ref{fig:rl:pl10k1.8} and \ref{fig:rl:er10k0.02}-\ref{fig:rl:er10k0.1}, \BaseNetPlus can achieve better results than \BaseNet in graphs with high density since the locality of vertices is significantly skew on the graph. 
On the other hand, for the graph with low density and flat degree distribution, such as Cora~(Fig.~\ref{fig:rl:cora}) and p2p~(Fig.~\ref{fig:rl:p2p}), using RL does not make significant improvement during the training.
We conjecture, in this scenario, the potential action space is relatively large so that it is difficult for RL algorithm to find an action trajectory with high reward, especially when the locality has been well captured by \BaseNet. 
In addition, due to the dynamic adjusting mechanism, \BaseNetPlus can make the training process robust and avoid over-fitting. As shown in Fig.~\ref{fig:rl:p2p} and \ref{fig:rl:ht}, \BaseNet reaches its best effect much earlier than \BaseNetPlus then it goes worse as over-fitting. 
In contrast, \BaseNetPlus can rectify the over-fitting autonomously in Fig.~\ref{fig:rl:fb} and \ref{fig:rl:pl10k2.0}. The performance of \BaseNetPlus increases constantly and surpasses \BaseNet in the end. 

\begin{figure}[t]
\includegraphics[width=0.7\linewidth]{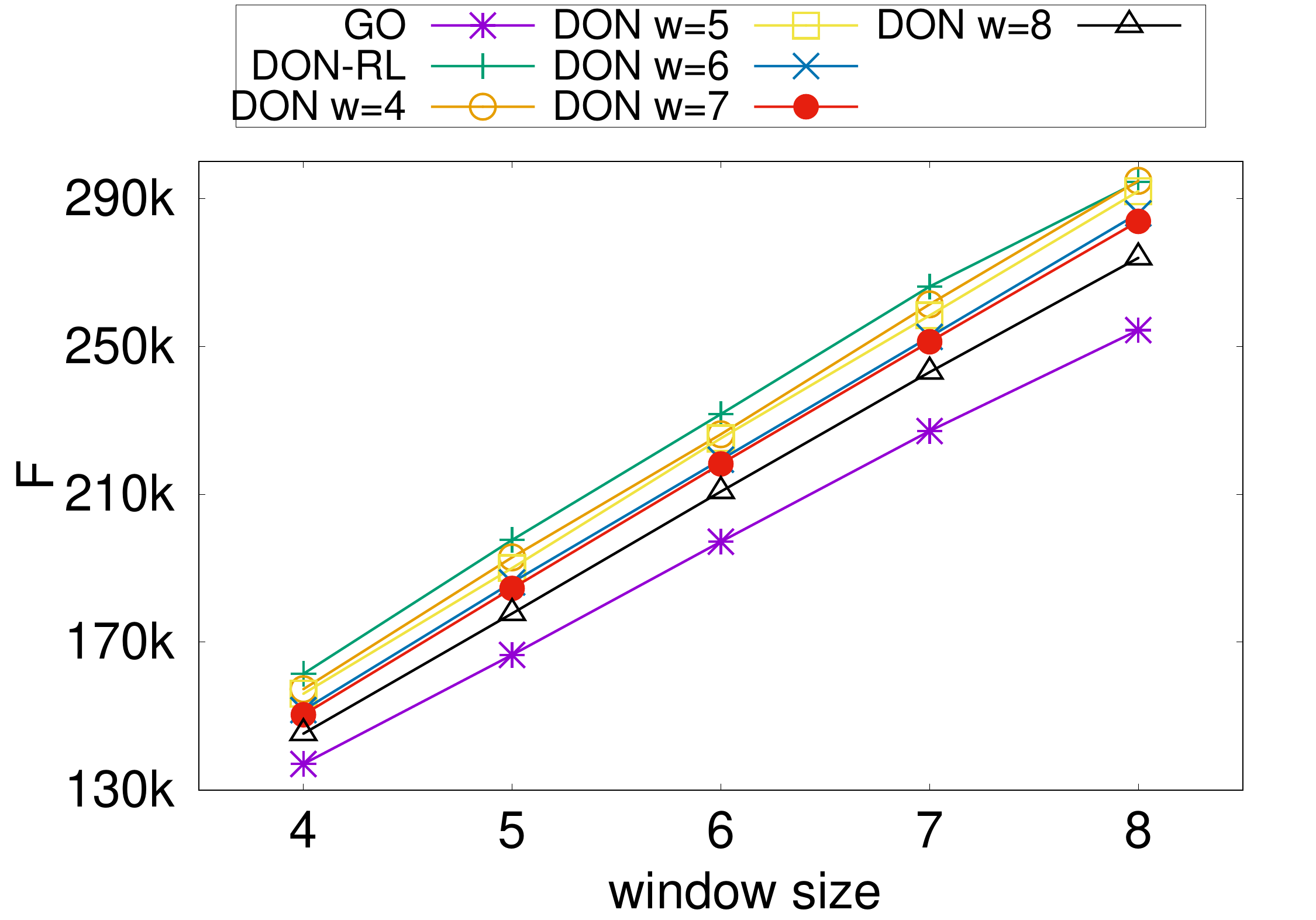}
\vspace*{-0.1cm}
\caption{Varying Window Size $w$}
\vspace*{-0.6cm}
\label{fig:window}
\end{figure}








Therefore, the RL-based \BaseNetPlus is suitable for the graphs whose structures are highly skew. The results in Fig.~\ref{fig:rl} also confirms the effectiveness of the Policy Network in \BaseNetPlus. 

\begin{figure*}[t]
\centering
\subfigure[ b = 8 ]{%
 \includegraphics[width=0.6\columnwidth]{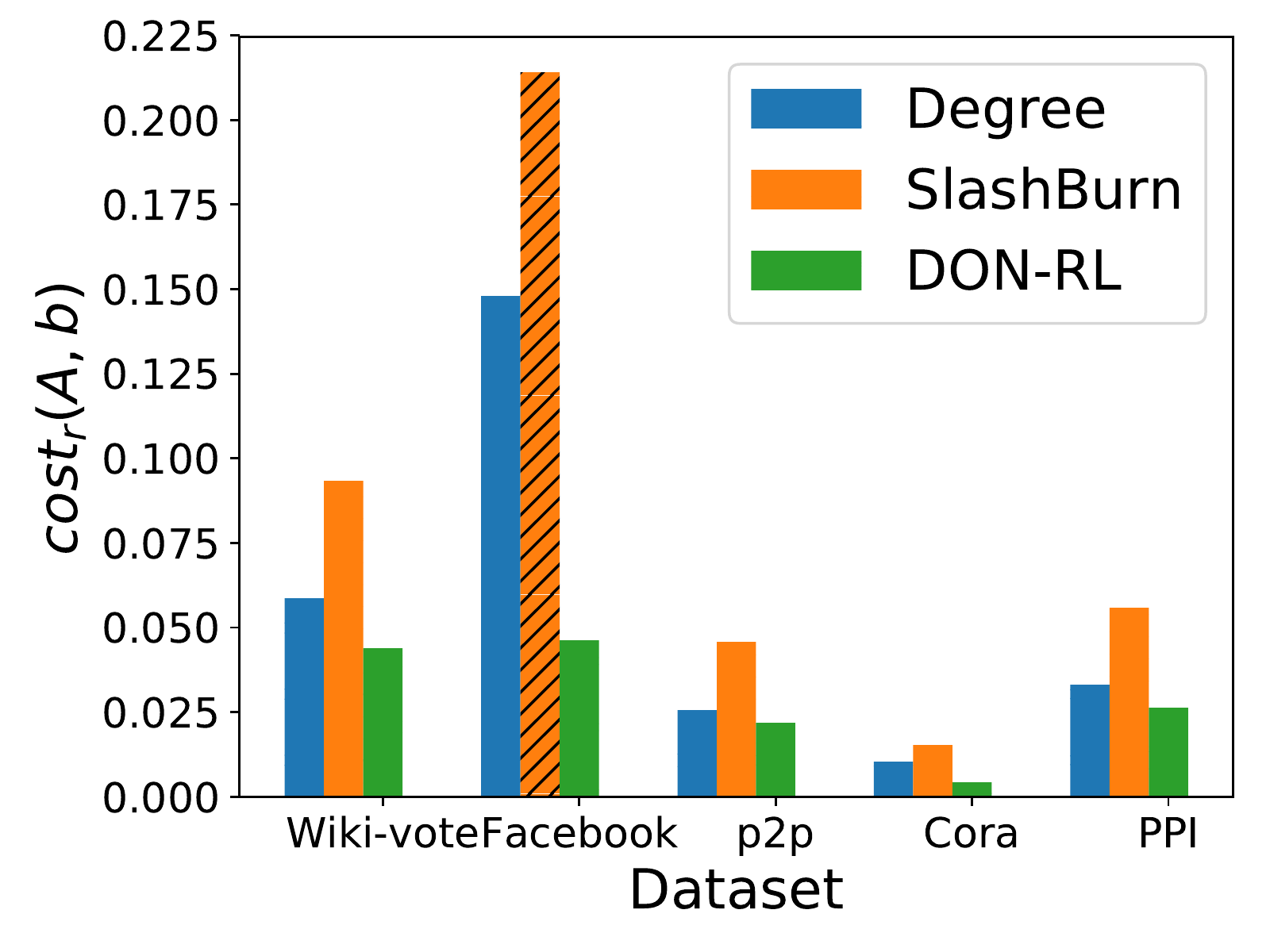} 
  \label{fig:compress:8}}
\subfigure[ b = 16 ]{%
  \includegraphics[width=0.6\columnwidth]{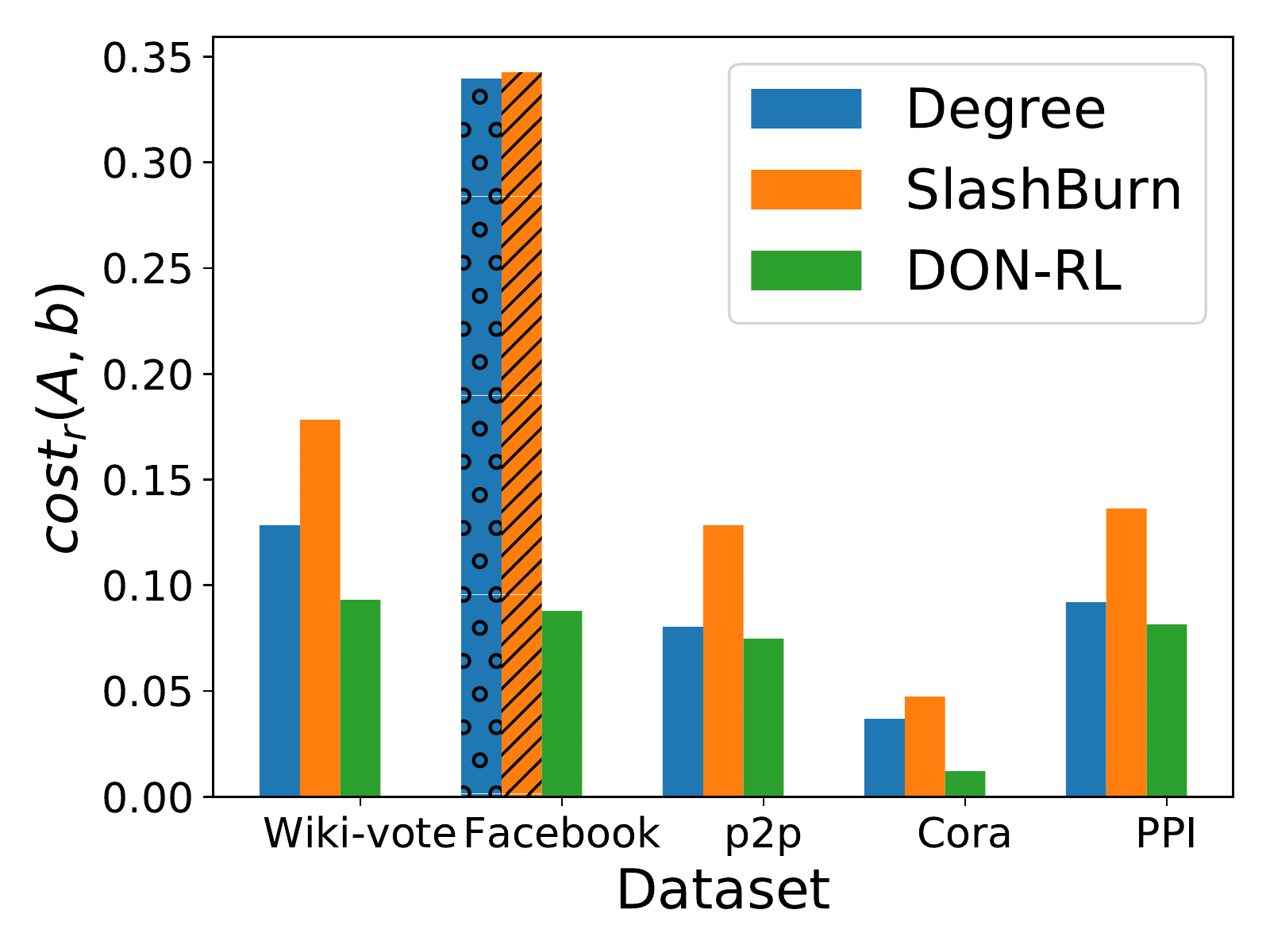}
  \label{fig:compress:16}}
 \subfigure[ b = 32 ]{%
  \includegraphics[width=0.6\columnwidth]{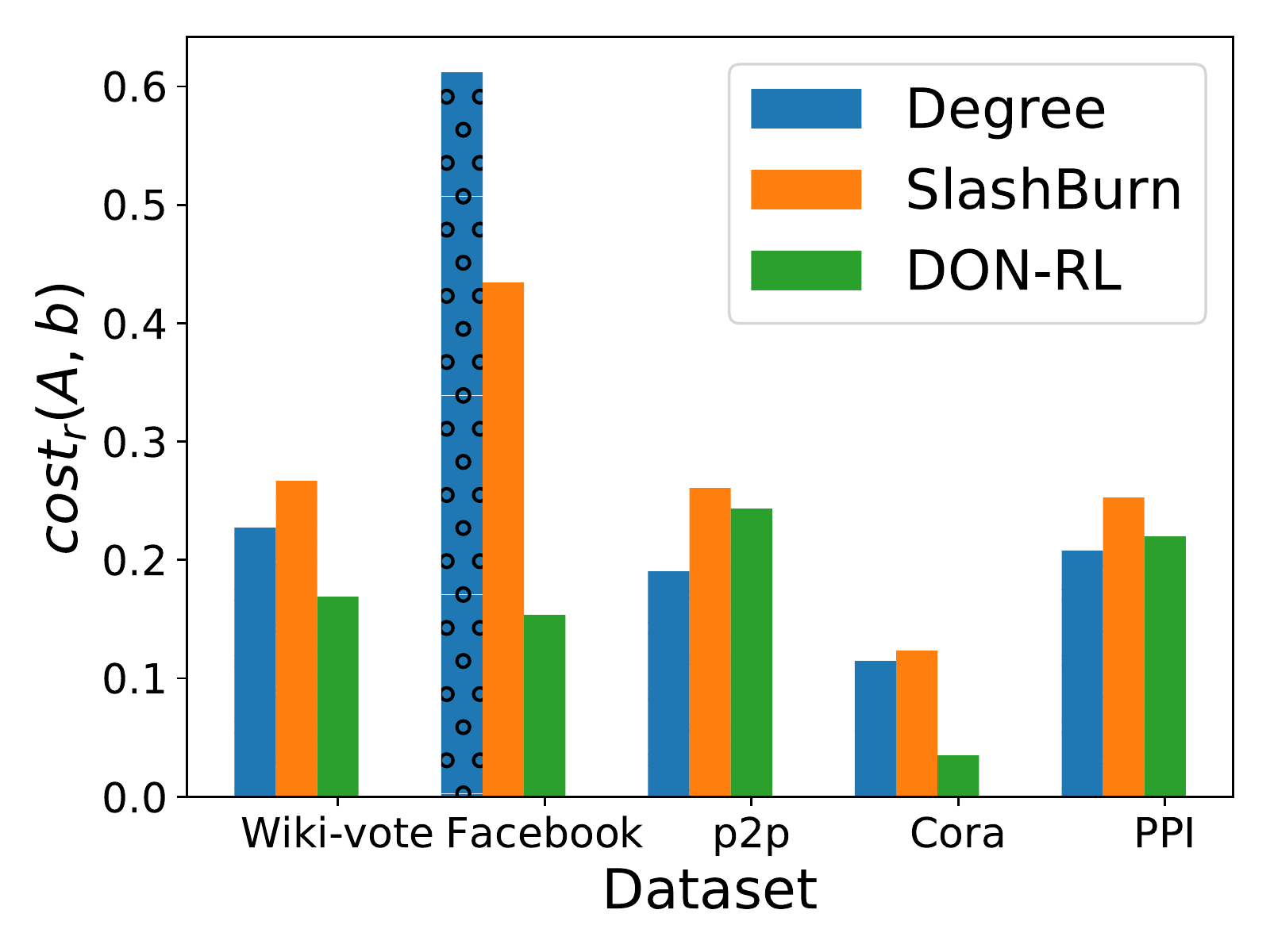}
  \label{fig:compress:32}}
\vspace*{-0.4cm}
\caption{Compression Costs}
\vspace*{-0.2cm}
\label{fig:casestudy2}
\end{figure*}

\begin{figure}[t]
\includegraphics[width=0.65\linewidth]{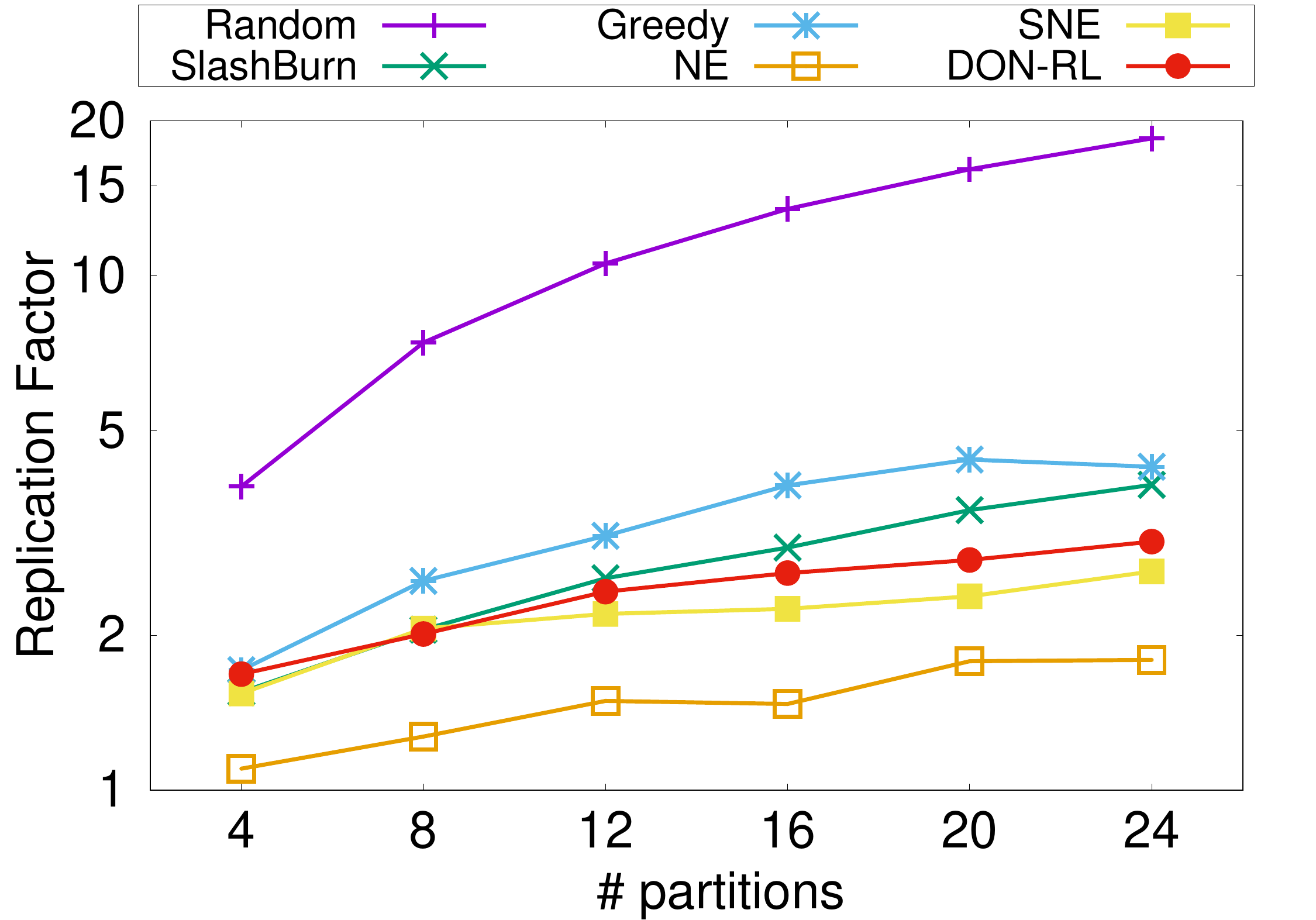}
\vspace*{-0.4cm}
\caption{Edge Partitioning of Facebook} 
\vspace*{-0.6cm}
\label{fig:casestudy}
\end{figure}

\subsection{Varying the Window Size $w$}
In this section, we investigate the effect of the window size $w$ in Eq.~(\ref{eq:fscore}). Fig.~\ref{fig:window} shows for graph PL10K\_1.6, the $F$ scores of 5 models training with $w$ is set to $\{4,5,6,7,8\}$. 
For these \BaseNet models, they perform stably better than \Gorder as $w$ grows.
In Fig.~\ref{fig:window}, \BaseNetPlus is the 5 models associated with RL training, which generates permutations of highest $F$.  
Recall that the approximate ratio of \Gorder is $\frac{1}{2w}$, which means theoretically, the larger the $w$, the larger the gap between \Gorder and the optimal solution, i.e., the improvement space of \Gorder. 
Interestingly, in Fig~\ref{fig:window}, as $w$ grows, the increment of locality score of model-based approaches grow a little faster than that of \Gorder. It implies that the model-based approaches can better take advantage of the gap to optimized the learned solution.

In addition, another interesting observation is that models trained with smaller $w$ have better performance than that of larger $w$. It indicates that we can use pre-trained models with small $w$ to generate different permutations of larger $w$.

\comments{
\begin{figure}[t]
\centering
    \includegraphics[width=0.4\textwidth]{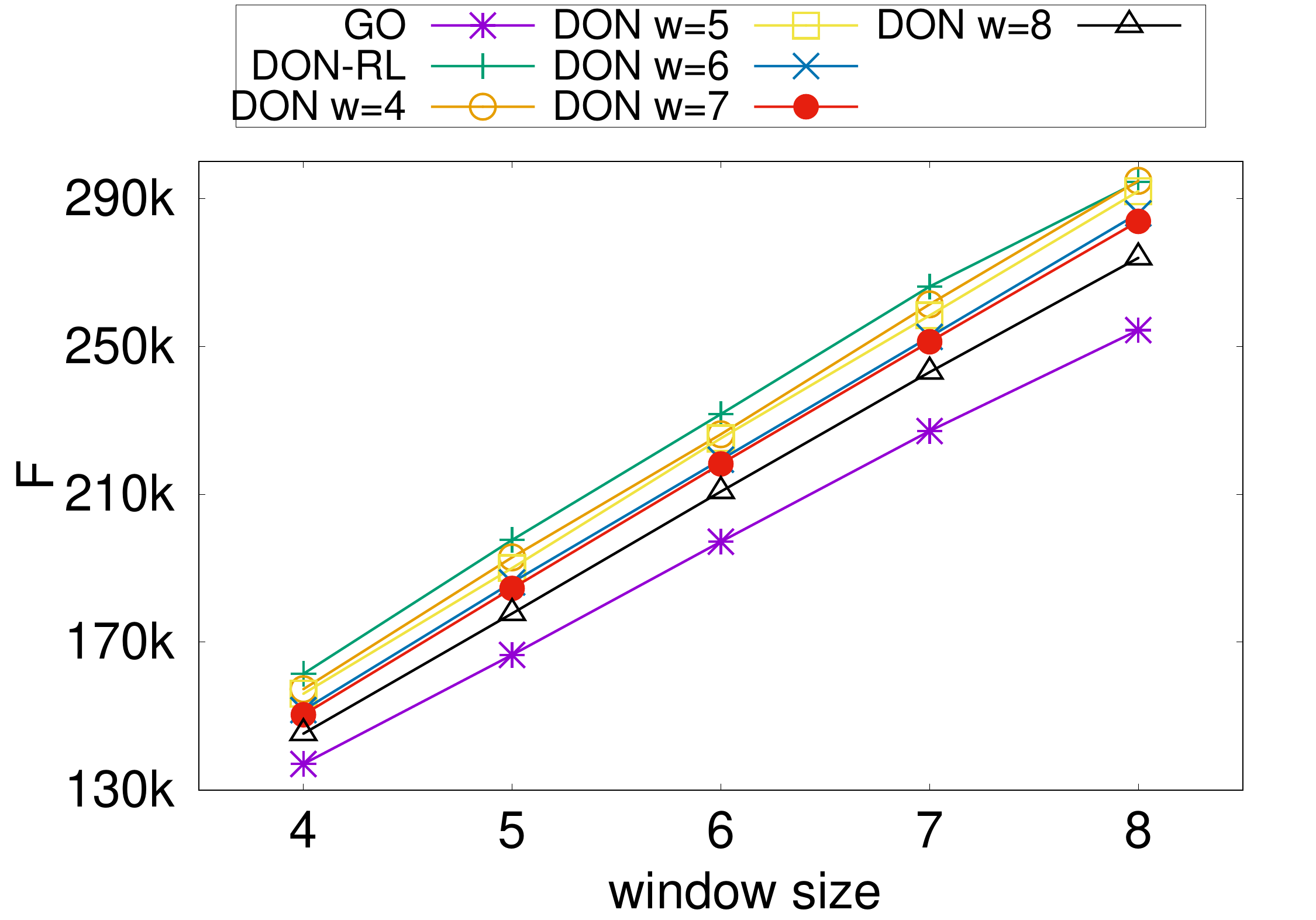}
    \vspace*{-0.2cm}
  \caption{Varying Window Size $w$}
   \label{fig:window}
\end{figure}
}

\subsection{Case Studies}
In this section, we perform two case studies to demonstrate the power of \BaseNetPlus in solving real-world problems. 

\stitle{Graph Compression:}
First, we explore the potential of using \BaseNetPlus to deal with \emph{graph compression} problem.
Concretely, given a graph, we want to layout its edges so that its adjacency matrix $A$ is easy to be compressed. 
Formally, this problem is equivalent to find a vertex ordering to minimize the storage cost $\COST_{nz}(A, b)$, where $b$ is the block width. 
Concretely, we divide the matrix $A$ into $b \times b$ square matrix blocks and count the number of the nonempty block as $\COST_{nz}(A, b)$ \cite{KangF11}. 
Since $\COST_{nz}(A, b)$ is not comparable for different block size $b$. Here we adopt a normalized version $\COST_{r}(A, b)$:
\begin{align}
\COST_{r}(A, b) = \frac{\COST_{nz}(A, b)}{\lceil N/b \rceil^2},
\end{align}
where $\lceil N/b \rceil^2$ is the number of blocks in $A$. A good ordering for the compression should have low $\COST_{nz}(A, b)$ for given block size $b$.

Fig.~\ref{fig:casestudy2} demonstrates the compression cost $\COST_r(A,b)$ on the 5 real graphs in Table~\ref{tbl:datasets}, when the block wide $b$ is $8$~(Fig.~\ref{fig:compress:8}),  $16$~(Fig.~\ref{fig:compress:16}) and $32$~(Fig.~\ref{fig:compress:32}).
We compare the compression performance with two methods: \Degree permutes the vertex based on the decreasing degree. \SlashBurn~\cite{KangF11} is the graph ordering algorithm proposed for graph compression. We set $w=5$ for our method: \BaseNetPlus for all datasets. As shown in Fig.~\ref{fig:casestudy2}, \BaseNetPlus outperforms the other methods on three datasets (Wiki\-vote, Facebook and Cora) and achieves comparable results on two datasets (p2p and PPI). For the number of nonempty blocks, \BaseNetPlus reduces the counts by $2.97\times$ and $2.25\times$ compared with the second best orderings on Facebook and Cora. Interestingly, \Degree is a quite good heuristic ordering for the graph compression. It achieves the best performance on p2p and PPI. The results validate that our learned vertex order is a good potential criteria for graph compression.

\stitle{Graph Edge Partitioning:}
The second application we investigate is \emph{graph edge partitioning}, which can achieve better load balancing than traditional vertex partitioning for skew graph data in the parallel/distribution environment. 
The problem is, given a partition number $k$, partitioning the edge set $E$ disjointly into $k$ balanced subsets $E_{i}$, to minimize the \emph{replication factor}~(RF) of the vertices in $k$ partitions:
\begin{equation}
\text{RF}(E_1,\cdots, E_k) = \frac{1}{|V|}\Sigma_{i \in [k]} |V(E_i)|, \text{where}  \cup_{i \in [k]}E_{i} = E,
\end{equation}
The problem is also NP-hardness~\cite{DBLP:conf/kdd/ZhangWLTL17}. Since the graph ordering is based on vertex locality, we use the ordering result of \BaseNetPlus to generate an edge partitioning directly. 
The partition is conducted by assigning edges of adjacent vertices in the permutation to one partition under the balance constraint. 

Fig~\ref{fig:casestudy} shows the partitioning derives from \BaseNetPlus with $w = 5$, and 5 other edge partitioning and graph ordering algorithms over Facebook. 
Here, \Random~\cite{DBLP:conf/kdd/BourseLV14} assigns each edge independently and uniformly to k partitions. 
The \Greedy~\cite{DBLP:conf/kdd/BourseLV14} heuristic preferentially assigns an edge to a partition already contains the two or one of its end vertices with trading off the balance. 
\NE and \SNE~\cite{DBLP:conf/kdd/ZhangWLTL17} are the state-of-the-art edge partitioning heuristic based on neighbor expansion. 
The replication factor of \BaseNetPlus is smaller than that of the widely used \Greedy up to 37\%. 
Also, as the number of partition increases, the replication factor of \BaseNetPlus increases much slower than \Greedy and \SlashBurn.
It is interesting that a good partitioning derives from a high-quality ordering result. 

\eat{
\begin{table*}[t]
{
\centering
 \caption{Maximum k Cover on Synthetic Graphs ($|V| = 1000$)}
\begin{tabular}{ |c|r|r|r|r|r|r|r|r|} \hline
  \multirow{2}{*}{\bf Dataset} & \multicolumn{2}{|c|}{\bf k = 5}  & \multicolumn{2}{|c|}{\bf k = 20 } & \multicolumn{2}{|c|}{\bf k = 100 }  & \multicolumn{2}{|c|}{\bf k = 300} \\\cline{2-9} 
  \multicolumn{1}{|c|}{} & Greedy  & {\sl BaseNet} & Greedy  & {\sl BaseNet} & Greedy &  {\sl BaseNet} &  Greedy  & {\sl BaseNet} \\ \cline{2-9}

 \hline 
 \hline
   0.01  & 95 & 89 & 353  & 332  & 1,492  & 14,22 & 3,473  & 3,309           \\ \hline
   0.05  & 356 & 337 & 1,330 & 1,295 & 5,872  & 5,660 & 1,4641 &  14,267        \\ \hline 			
   0.1   & 638 & 609 & 2,453 & 2,360  & 11,185 & 10,907 & 28,424 &  27,793       \\ \hline 
   0.2   & 1,173 & 1,156 & 4,562 & 4,466  & 21,146 & 20,755  & 54,681 & 53,700    \\ \hline 
   0.4   & 2,191 & 2,142 & 8,629 & 8,500  & 40,582  & 39,958  & 106,506  & 105,045   \\ \hline 
   0.8   & 4,151 & 4,110 & 16,395 & 16,213 & 77,952 & 77,256  & 207,363 & 205,965   \\ \hline 
 
    \end{tabular}
      \vspace*{0.2cm}

\label{tbl:sql2graph}
}
%
\end{table*}
}

\section{Related Works}  
\label{sec:rw} 
\stitle{Graph Representation Learning} is to learn 
%
%
an effective graph representation to encode sufficient features for
downstream machine learning/deep learning tasks.
%
%
Recently, graph representation learning have been extensively
studied. The surveys can be found in \cite{cui2018survey,
  DBLP:journals/debu/HamiltonYL17}.
The traditional graph representation learning, that uses matrix
factorization to find a low-rank space for the adjacency
matrix~\cite{DBLP:conf/www/AhmedSNJS13, DBLP:conf/nips/BelkinN01,
  fu2012graph}, cannot be used to solve our problem, because they are designed to reconstruct the original graph.
%
%
Structure preserving representation learning techniques can
encode/decode graph structures and properties. \DeepWalk
~\cite{DBLP:conf/kdd/PerozziAS14} and
\NodetoVec~\cite{DBLP:conf/kdd/GroverL16} encode the neighborhood
relationship by exploiting truncated random walks as embedding
context.
%
%
Furthermore, there are the graph representations to preserve vertex
information at different granularity from microscopic neighborhood
structure to macroscopic community structure.
\cite{DBLP:conf/www/TangQWZYM15, DBLP:conf/www/TsitsulinMKM18,
  DBLP:conf/kdd/ZhangCWPY018} carry high-order proximity and closeness
measures, to provide flexible and diverse representations for
unsupervised graph learning tasks.
\MNMF~\cite{DBLP:conf/aaai/WangCWP0Y17} incorporates graph macroscopic
proximity, the community affiliation of vertex.
Such approaches are designed to machine learning tasks (e.g., node
classification, link prediction, as well as recommendation).
%
%
Unfortunately, these representations cannot be directly applied to our
problem. First, these representations are learned by unsupervised
fashion.  Second, they do not contain abundant vertex features, the
learned graph proximity is insufficient.  Third, our problem aims to
find an optimal combinatorial structure instead of performing
inference on individual instances which only involves local
information.

There are reported studies on representation learning over special graphs, such
as representation of heterogeneous
graphs~\cite{DBLP:conf/kdd/DongCS17}, dynamic
graphs \cite{DBLP:journals/corr/abs-1803-04051}, attributed
graphs~\cite{DBLP:journals/corr/KipfW16}.
%
%
And there are specified graph machine learning applications that
require specialized graph representation, for instance, influence
diffusion prediction \cite{DBLP:conf/icde/FengCKLLC18}, anomaly
detection~\cite{DBLP:conf/icde/HuAMH16}, and network
alignment~\cite{DBLP:conf/ijcai/ManSLJC16}.  These models are too
specific to deal with our problem.

\stitle{Neural Networks for Combinatorial Optimization}: Early works
that use neural network to construct solutions for NP-hard
optimization problems are summarized in
\cite{DBLP:journals/informs/Smith99}.
Recently, deep learning has also been adopted to solve combinatorial
optimization. A new attention-based sequence-to-sequence neural
architecture, Pointer Network~\cite{NIPS2015_5866}, is proposed and is
used to solve the planar Travel Salesman Problem~(TSP). Pointer
Network learns the planar convex hull supervised by a set of problem
instances and solutions alone,
and then \cite{DBLP:journals/corr/BelloPLNB16} uses reinforcement
learning, the Actor-Critic algorithm~\cite{DBLP:conf/nips/KondaT99},
to train Pointer Network. Instead of using given training instances,
an actor network makes trials of tour and evaluates
the trials using a critic network.
%
To further improve the performance, \cite{anonymous2019attention}
proposes an alternative neural network which encodes input nodes with
multi-head attention~\cite{DBLP:conf/nips/VaswaniSPUJGKP17}.
These
approaches~\cite{NIPS2015_5866,DBLP:journals/corr/BelloPLNB16,anonymous2019attention}
concentrate on 2D Euclidean space TSP, and it
is non-trivial to extend them to deal with graphs.
For graph data, 
%
Dai et. al.~\cite{DBLP:conf/nips/KhalilDZDS17} propose a framework to
learn a greedy meta-algorithm over graphs. First, a graph is encoded
by a graph embedding network. Second, the embedding is fed into some
more layers to estimate an evaluation function.  The overall network
is trained by deep Q-learning~\cite{DBLP:journals/ml/Williams92} in an
end-to-end fashion.  The framework is demonstrated to solve Minimum
Vertex Cover, Maximum Cut and TSP.  These studies aim to generalize
the process of problem-solving for a distribution of problem instances
offline.
%
A recent study provides a
supervised learning approach to deal with NP problems equivalent to
maximum independent set (MIS).  It adopts graph convolutional network
(GCN)~\cite{DBLP:journals/corr/KipfW16} to predict the likelihood whether each
vertex is in the optimal solution~\cite{DBLP:conf/nips/LiCK18}.
Since it relies on the state-of-the-art MIS local search heuristics
to refine the candidate solutions, the bare utility of the model needs
to be further studied.
%
%

\section{Conclusion}
\label{sec:conclusion}
In this paper, we focus on an NP-hard problem, graph
ordering, in a novel, machine learning-based perspective.
Distinguished from recent research, the NP-hard problem is over an
 specific larger graph. We design a new model: Deep Order Network~(\BaseNet)
to learn the underlying combinatorial closeness over the
vertices of graph, by sampling small vertex sets as locally partial
solutions. To further improve the sampling effectiveness, we propose an
adaptive training approach based on reinforcement learning, which
automatically adjusts the sampling probabilities. 
Compare with the high-quality greedy algorithm, our overall model, \BaseNetPlus 
improves the quality of solution up to 7.6\% and 35\% for real graphs and
synthetic graphs, respectively.  
Our study reveals that a simple neural network has the
ability to deal with NP optimization problem by encoding hidden features of the combination structures.
We will public our code and pre-trained models later.

{
\bibliographystyle{abbrv}
\bibliography{ref}

\begin{thebibliography}{10}

\bibitem{konect}
{KONECT} (the koblenz network collection).
\newblock \url{http://konect.uni-koblenz.de}.

\bibitem{DBLP:conf/osdi/AbadiBCCDDDGIIK16}
M.~Abadi, P.~Barham, J.~Chen, Z.~Chen, A.~Davis, J.~Dean, M.~Devin,
  S.~Ghemawat, G.~Irving, M.~Isard, M.~Kudlur, J.~Levenberg, R.~Monga,
  S.~Moore, D.~G. Murray, B.~Steiner, P.~A. Tucker, V.~Vasudevan, P.~Warden,
  M.~Wicke, Y.~Yu, and X.~Zheng.
\newblock Tensorflow: {A} system for large-scale machine learning.
\newblock In {\em Proc. of {OSDI}'16}, 2016.

\bibitem{DBLP:conf/www/AhmedSNJS13}
A.~Ahmed, N.~Shervashidze, S.~M. Narayanamurthy, V.~Josifovski, and A.~J.
  Smola.
\newblock Distributed large-scale natural graph factorization.
\newblock In {\em Proc. of {WWW}'13}, 2013.

\bibitem{anonymous2019attention}
Anonymous.
\newblock Attention, learn to solve routing problems!
\newblock In {\em Submitted to International Conference on Learning
  Representations}, 2019.

\bibitem{applegate2006concorde}
D.~Applegate, R.~Bixby, V.~Chvatal, and W.~Cook.
\newblock Concorde tsp solver, 2006.

\bibitem{DBLP:journals/cgf/BehrischBRSF16}
M.~Behrisch, B.~Bach, N.~H. Riche, T.~Schreck, and J.~Fekete.
\newblock Matrix reordering methods for table and network visualization.
\newblock {\em Comput. Graph. Forum}, 35(3), 2016.

\bibitem{DBLP:conf/nips/BelkinN01}
M.~Belkin and P.~Niyogi.
\newblock Laplacian eigenmaps and spectral techniques for embedding and
  clustering.
\newblock In {\em Proc. of {NIPS}'01}, 2001.

\bibitem{DBLP:journals/corr/BelloPLNB16}
I.~Bello, H.~Pham, Q.~V. Le, M.~Norouzi, and S.~Bengio.
\newblock Neural combinatorial optimization with reinforcement learning.
\newblock {\em CoRR}, abs/1611.09940, 2016.

\bibitem{DBLP:conf/kdd/BourseLV14}
F.~Bourse, M.~Lelarge, and M.~Vojnovic.
\newblock Balanced graph edge partition.
\newblock In {\em Proc. of {KDD}'14}, 2014.

\bibitem{cplex201412}
I.~I. CPLEX.
\newblock 12.6.
\newblock {\em CPLEX User?s Manual}, 2014.

\bibitem{csaji2001approximation}
B.~C. Cs{\'a}ji.
\newblock Approximation with artificial neural networks.
\newblock {\em Faculty of Sciences, Etvs Lornd University, Hungary}, 24:48,
  2001.

\bibitem{cui2018survey}
P.~Cui, X.~Wang, J.~Pei, and W.~Zhu.
\newblock A survey on network embedding.
\newblock {\em IEEE Trans. on Knowledge and Data Engineering}, 2018.

\bibitem{DBLP:conf/kdd/DongCS17}
Y.~Dong, N.~V. Chawla, and A.~Swami.
\newblock metapath2vec: Scalable representation learning for heterogeneous
  networks.
\newblock In {\em Proc. of {KDD}'17}, 2017.

\bibitem{DBLP:conf/icde/FengCKLLC18}
S.~Feng, G.~Cong, A.~Khan, X.~Li, Y.~Liu, and Y.~M. Chee.
\newblock Inf2vec: Latent representation model for social influence embedding.
\newblock In {\em Proc. of {ICDE}'18}, 2018.

\bibitem{fu2012graph}
Y.~Fu and Y.~Ma.
\newblock {\em Graph embedding for pattern analysis}.
\newblock Springer Science \& Business Media, 2012.

\bibitem{graphlab}
J.~E. Gonzalez, Y.~Low, H.~Gu, D.~Bickson, and C.~Guestrin.
\newblock Powergraph: Distributed graph-parallel computation on natural graphs.
\newblock In {\em Proc. of OSDI'12}, 2012.

\bibitem{DBLP:conf/osdi/GonzalezXDCFS14}
J.~E. Gonzalez, R.~S. Xin, A.~Dave, D.~Crankshaw, M.~J. Franklin, and
  I.~Stoica.
\newblock Graphx: Graph processing in a distributed dataflow framework.
\newblock In {\em Proc. of OSDI'14}, 2014.

\bibitem{DBLP:conf/icml/GopalanMM14}
A.~Gopalan, S.~Mannor, and Y.~Mansour.
\newblock Thompson sampling for complex online problems.
\newblock In {\em Proc. of {ICML}'14}, 2014.

\bibitem{DBLP:conf/kdd/GroverL16}
A.~Grover and J.~Leskovec.
\newblock node2vec: Scalable feature learning for networks.
\newblock In {\em Proc. of {KDD}'16}, 2016.

\bibitem{DBLP:journals/debu/HamiltonYL17}
W.~L. Hamilton, R.~Ying, and J.~Leskovec.
\newblock Representation learning on graphs: Methods and applications.
\newblock {\em {IEEE} Data Eng. Bull.}, 40(3), 2017.

\bibitem{DBLP:conf/icde/HuAMH16}
R.~Hu, C.~C. Aggarwal, S.~Ma, and J.~Huai.
\newblock An embedding approach to anomaly detection.
\newblock In {\em Proc. of {ICDE}'16}, 2016.

\bibitem{KangF11}
U.~Kang and C.~Faloutsos.
\newblock Beyond 'caveman communities': Hubs and spokes for graph compression
  and mining.
\newblock In {\em Proc. of {ICDM}'11}, 2011.

\bibitem{DBLP:conf/nips/KhalilDZDS17}
E.~B. Khalil, H.~Dai, Y.~Zhang, B.~Dilkina, and L.~Song.
\newblock Learning combinatorial optimization algorithms over graphs.
\newblock In {\em Proc. of {NIPS}'17}, 2017.

\bibitem{DBLP:journals/corr/KingmaB14}
D.~P. Kingma and J.~Ba.
\newblock Adam: {A} method for stochastic optimization.
\newblock In {\em Proc. of ICLR'15}, 2015.

\bibitem{DBLP:journals/corr/KipfW16}
T.~N. Kipf and M.~Welling.
\newblock Semi-supervised classification with graph convolutional networks.
\newblock {\em CoRR}, abs/1609.02907, 2016.

\bibitem{DBLP:conf/nips/KondaT99}
V.~R. Konda and J.~N. Tsitsiklis.
\newblock Actor-critic algorithms.
\newblock In {\em Proc. of {NIPS}'99}, 1999.

\bibitem{snapnets}
J.~Leskovec and A.~Krevl.
\newblock {SNAP Datasets}: {Stanford} large network dataset collection.
\newblock \url{http://snap.stanford.edu/data}, June 2014.

\bibitem{DBLP:conf/nips/LiCK18}
Z.~Li, Q.~Chen, and V.~Koltun.
\newblock Combinatorial optimization with graph convolutional networks and
  guided tree search.
\newblock In {\em Proc. of {NeurIPS}'18}, 2018.

\bibitem{DBLP:conf/ijcai/ManSLJC16}
T.~Man, H.~Shen, S.~Liu, X.~Jin, and X.~Cheng.
\newblock Predict anchor links across social networks via an embedding
  approach.
\newblock In {\em Proc. of {IJCAI}'16}, 2016.

\bibitem{newman2001random}
M.~E. Newman, S.~H. Strogatz, and D.~J. Watts.
\newblock Random graphs with arbitrary degree distributions and their
  applications.
\newblock {\em Physical review E}, 64(2):026118, 2001.

\bibitem{DBLP:conf/kdd/PerozziAS14}
B.~Perozzi, R.~Al{-}Rfou, and S.~Skiena.
\newblock Deepwalk: online learning of social representations.
\newblock In {\em Proc. of {KDD}'14}, 2014.

\bibitem{DBLP:journals/nature/SilverHMGSDSAPL16}
D.~Silver, A.~Huang, C.~J. Maddison, A.~Guez, L.~Sifre, G.~van~den Driessche,
  J.~Schrittwieser, I.~Antonoglou, V.~Panneershelvam, M.~Lanctot, S.~Dieleman,
  D.~Grewe, J.~Nham, N.~Kalchbrenner, I.~Sutskever, T.~P. Lillicrap, M.~Leach,
  K.~Kavukcuoglu, T.~Graepel, and D.~Hassabis.
\newblock Mastering the game of go with deep neural networks and tree search.
\newblock {\em Nature}, 529(7587), 2016.

\bibitem{DBLP:journals/informs/Smith99}
K.~A. Smith.
\newblock Neural networks for combinatorial optimization: {A} review of more
  than a decade of research.
\newblock {\em {INFORMS} Journal on Computing}, 11(1), 1999.

\bibitem{DBLP:books/lib/SuttonB98}
R.~S. Sutton and A.~G. Barto.
\newblock {\em Reinforcement learning - an introduction}.
\newblock Adaptive computation and machine learning. {MIT} Press, 1998.

\bibitem{DBLP:conf/www/TangQWZYM15}
J.~Tang, M.~Qu, M.~Wang, M.~Zhang, J.~Yan, and Q.~Mei.
\newblock {LINE:} large-scale information network embedding.
\newblock In {\em Proc. of {WWW}'15}, 2015.

\bibitem{DBLP:journals/corr/abs-1803-04051}
R.~Trivedi, M.~Farajtabar, P.~Biswal, and H.~Zha.
\newblock Representation learning over dynamic graphs.
\newblock {\em CoRR}, abs/1803.04051, 2018.

\bibitem{DBLP:conf/www/TsitsulinMKM18}
A.~Tsitsulin, D.~Mottin, P.~Karras, and E.~M{\"{u}}ller.
\newblock {VERSE:} versatile graph embeddings from similarity measures.
\newblock In {\em Proc. of {WWW}'18}, 2018.

\bibitem{DBLP:journals/jdwm/TsoumakasK07}
G.~Tsoumakas and I.~Katakis.
\newblock Multi-label classification: An overview.
\newblock {\em {IJDWM}}, 3(3), 2007.

\bibitem{DBLP:conf/nips/VaswaniSPUJGKP17}
A.~Vaswani, N.~Shazeer, N.~Parmar, J.~Uszkoreit, L.~Jones, A.~N. Gomez,
  L.~Kaiser, and I.~Polosukhin.
\newblock Attention is all you need.
\newblock In {\em Proc. of {NIPS}'17}, 2017.

\bibitem{NIPS2015_5866}
O.~Vinyals, M.~Fortunato, and N.~Jaitly.
\newblock Pointer networks.
\newblock In C.~Cortes, N.~D. Lawrence, D.~D. Lee, M.~Sugiyama, and R.~Garnett,
  editors, {\em Advances in Neural Information Processing Systems 28}. 2015.

\bibitem{DBLP:conf/aaai/WangCWP0Y17}
X.~Wang, P.~Cui, J.~Wang, J.~Pei, W.~Zhu, and S.~Yang.
\newblock Community preserving network embedding.
\newblock In {\em Proc. of {AAAI}'17}, 2017.

\bibitem{DBLP:conf/sigmod/WeiYLL16}
H.~Wei, J.~X. Yu, C.~Lu, and X.~Lin.
\newblock Speedup graph processing by graph ordering.
\newblock In {\em Proc. of {SIGMOD}'16}, 2016.

\bibitem{DBLP:journals/ml/Williams92}
R.~J. Williams.
\newblock Simple statistical gradient-following algorithms for connectionist
  reinforcement learning.
\newblock {\em Machine Learning}, 8, 1992.

\bibitem{DBLP:conf/nips/ZaheerKRPSS17}
M.~Zaheer, S.~Kottur, S.~Ravanbakhsh, B.~P{\'{o}}czos, R.~R. Salakhutdinov, and
  A.~J. Smola.
\newblock Deep sets.
\newblock In {\em Proc. of {NIPS}'17}, 2017.

\bibitem{DBLP:conf/kdd/ZhangWLTL17}
C.~Zhang, F.~Wei, Q.~Liu, Z.~G. Tang, and Z.~Li.
\newblock Graph edge partitioning via neighborhood heuristic.
\newblock In {\em Proc. of {KDD}'17}, 2017.

\bibitem{DBLP:conf/kdd/ZhangCWPY018}
Z.~Zhang, P.~Cui, X.~Wang, J.~Pei, X.~Yao, and W.~Zhu.
\newblock Arbitrary-order proximity preserved network embedding.
\newblock In {\em Proc. of {KDD}'18}, 2018.

\end{thebibliography}
}


\end{document}